\documentclass[aps,prb,twocolumn,superscriptaddress]{revtex4-1}

\usepackage{amsmath}
\usepackage{hyperref}
\usepackage{graphicx}
\usepackage{bbold} % for identity matrix
\usepackage{todonotes}

\newcommand{\Ham}{\mathcal{H}} %for the Hamiltonian H
\newcommand{\w}{\omega} %for the Greek letter omega
\newcommand{\diff}{\mathrm{d}} %differential d
\newcommand{\Ej}{E_{\mathrm{J}}} %Josepshon energy
\newcommand{\wpp}{\w_{\rm{p}}} %plasma frequency of one junction
\newcommand{\e}{\mathrm{e}} %exponential function
\newcommand{\BesselI}{\mathrm{I}}
\newcommand{\StruveL}{\mathrm{L}}
\newcommand{\ug}{u_{\rm g}}

\def\be{\begin{equation}}
\def\ee{\end{equation}}
\def\bea{\begin{eqnarray}}
\def\eea{\end{eqnarray}}

\begin{document}

\title{Superconductor-Insulator Transition in disordered Josephson junction chains}

\author{M. Bard}
\affiliation{Institut f\"ur Nanotechnologie, Karlsruhe Institute of Technology, 76021 Karlsruhe, Germany}

\author{I.~V. Protopopov}
\affiliation{Department of Theoretical Physics, University of Geneva, 1211 Geneva, Switzerland  }
\affiliation{Institut f\"ur Nanotechnologie, Karlsruhe Institute of Technology, 76021 Karlsruhe, Germany}
\affiliation{Landau Institute for Theoretical Physics, 119334 Moscow, Russia}

\author{I.~V. Gornyi}
\affiliation{Institut f\"ur Nanotechnologie, Karlsruhe Institute of Technology, 76021 Karlsruhe, Germany}
\affiliation{Institut f\"ur Theorie der Kondensierten Materie, Karlsruhe Institute of Technology, 76128 Karlsruhe, Germany} 
\affiliation{Landau Institute for Theoretical Physics, 119334 Moscow, Russia}
\affiliation{A. F. Ioffe Physico-Technical Institute, 194021 St. Petersburg, Russia}

\author{A. Shnirman}
\affiliation{Institut f\"ur Theorie der Kondensierten Materie, Karlsruhe Institute of Technology, 76128 Karlsruhe, Germany} 
\affiliation{Landau Institute for Theoretical Physics, 119334 Moscow, Russia}

\author{A.~D. Mirlin}
\affiliation{Institut f\"ur Nanotechnologie, Karlsruhe Institute of Technology, 76021 Karlsruhe, Germany}
\affiliation{Institut f\"ur Theorie der Kondensierten Materie, Karlsruhe Institute of Technology, 76128 Karlsruhe, Germany} 
\affiliation{Landau Institute for Theoretical Physics, 119334 Moscow, Russia}
\affiliation{Petersburg Nuclear Physics Institute, 188350 St. Petersburg, Russia}

\date{\today}

\begin{abstract}
We study the superconductor-insulator quantum phase transition in disordered Josephson junction chains.
To this end, we derive the field theory from the lattice model that describes a chain of superconducting islands
with a capacitive coupling to the ground ($C_0$) as well as between
the islands ($C_1$). We analyze the theory in the
short-range ($C_1 \ll C_0$) and in the long-range ($C_1 \gg C_0$)  limits.
The transition to the insulating state is driven by the proliferation of quantum phase slips. The most important source of disorder originates from trapped charges in the substrate that suppress the coherence of phase slips, thus favoring superconducting correlations.
Using the renormalization-group approach, we determine the phase
diagram and evaluate the temperature dependence of the dc conductivity and system-size dependence of the resistance around the superconductor-insulator transition. These dependences have in general strongly non-monotonic character, with several distinct regimes reflecting an intricate interplay of superconductivity and disorder. 
\end{abstract}

% insert suggested PACS numbers in braces on next line
% insert suggested keywords - APS authors don't need to do this
%\keywords{}

%\maketitle must follow title, authors, abstract, \pacs, and \keywords
\maketitle

\section{Introduction} 
\label{Sec:Introduction}

One-dimensional (1D) Josephson junction (JJ) chains show a remarkably rich physics. In the insulating regime, the Coulomb blockade for Cooper-pair tunneling can be observed \cite{HavilandDelsing96}. This effect is characterized by a zero current state below a certain threshold voltage at zero temperature $T$.  At $T\neq 0$, thermally activated hopping of Cooper pairs has been observed \cite{ZimmerEtAl13}. Above the threshold voltage, transport is governed by charge solitons\cite{HermonEtAl96}, kink-like excitations that show relativistic effects like Lorentz contraction. Furthermore, in the case of strong charge disorder, depinning effects are the dominant mechanism for the onset of transport above the threshold voltage \cite{VogtEtAl15}.  In the conducting regime, where the Josephson energy dominates over the charging energy, the current-voltage curve shows a supercurrent-like behavior at low bias voltages and a constant current at higher voltages \cite{ErguelEtAl13,ErguelEtAl13b}. 

Another interesting effect is the persistent current that arises if a closed chain is pierced by a magnetic flux\cite{MatveevEtAl02,PopEtAl10,RastelliEtAl13}. In the classical regime of large Josephson coupling, a sawtooth-like shape of the current-phase relation is found, with a rounding near the transition points due to quantum fluctuations originating from a finite charging energy. In the limit of strong fluctuations, the relation develops a sinusoidal shape.  
The most important fluctuations leading to this behavior are quantum phase slips (QPS)---processes in which the phase difference across the ring changes by $2\pi$. Quantum phase slips are also vital for a number of recently suggested applications  of 1D JJ chains in the context of metrology \cite{GuichardHekking10} and decoherence-protected quantum computations \cite{DoucotEtAl02,ProtopopovFeigelman04,ProtopopovFeigelman06, DoucotEtAl05,GladchenkoEtAl09,BellEtAl14}.

%The most important fluctuations leading to this behavior are quantum phase slips (QPS). In the QPS process, the total phase difference across the ring changes by $2\pi$.
%Replacing simple junctions by Rhombi-shaped devices (four junctions) gives rise to tunneling of pairs of Cooper pairs which results in a $4e$-periodic supercurrent\cite{DoucotEtAl02,ProtopopovFeigelman04,ProtopopovFeigelman06}. A possible application for such devices is the construction of qubits protected against decoherence\cite{DoucotEtAl05,GladchenkoEtAl09,BellEtAl14}. A different application of JJ chains in general is a realization of the current standard in metrology \cite{GuichardHekking10}.

Using a SQUID geometry enables tuning of the Josephson energy in situ by applying a perpendicular magnetic field. This provides a convenient way for exploration of the superconductor-insulator transition (SIT) \cite{ChowEtAl98,HavilandEtAl00,KuoChen01,MiyazakiEtAl02,TakahideEtAl06}. An early theoretical description of JJ chains was introduced by Bradley and Doniach\cite{BradleyDoniach84} who considered a model with capacitive coupling to the ground that can be mapped onto a two-dimensional (2D) XY-model showing a Berezinskii-Kosterlitz-Thoulesss (BKT) transition\cite{Berezinskii71,Berezinskii72,KosterlitzThouless73}.  An alternative model, with junction capacitances only, was considered in Ref.~\onlinecite{Fisher88}. This model is characterized by insulating behavior independent of the Josephson energy since quantum fluctuations destroy phase coherence. In later works, the theory of Ref.~\onlinecite{BradleyDoniach84} has been extended by including dissipation\cite{Korshunov89,BobbertEtAl90,BobbertEtAl92} and considering both capacitive couplings (to the ground and between the islands)\cite{Korshunov89,ChoiEtAl98,RastelliEtAl13}. 
Also, a connection to the Luttinger-liquid physics has been pointed out\cite{Fazio96,Glazman97,Fazio01}.
Effects of disorder in JJ chains were studied in the context of persistent current\cite{MatveevEtAl02,ProtopopovFeigelman06}. It was found, in particular, that random offset charges destroy the coherence of QPS leading to a weaker decay of the amplitude of the supercurrent in chains with ring structure. One thus may expect that disorder should play an important role also for the physics of SIT in JJ chains.

The SIT is a remarkable quantum phase transition which separates two antagonist phases---the superconducting one with zero resistivity and the insulating one with infinite resistivity. The SIT arises naturally in low-dimensional systems, i.e., in 1D and 2D geometry, since the Anderson localization precludes, under conventional circumstances,  the emergence of an intermediate metallic phase. A particularly large body of work has been carried out on SIT in 2D geometry.  Specifically, such a transition was studied experimentally in a large variety of 2D structures and materials, including JJ arrays [\onlinecite{Geerligs89,Chen95}], amorphous Bi and Pb [\onlinecite{Haviland89,Parendo05}], MoC
[\onlinecite{Lee90}], MoGe [\onlinecite{Yazdani95}], Ta [\onlinecite{Qin06}], InO [\onlinecite{Exp-InO,Exp-InO-1,Exp-InO-2,Sherman14,Ovadia13,Ovadia15}], NbN [\onlinecite{Exp-NbN}] and TiN films [\onlinecite{Exp-TiN,BatVinBKT,Baturina13}],  LaAlO$_3$/SrTiO$_3$ interfaces
[\onlinecite{Caviglia2008,Ilani2014}], SrTiO$_3$ surfaces [\onlinecite{Kim2012,Iwasa2014}],
MoS$_2$ flakes [\onlinecite{Ye2012,Taniguchi2012}], FeSe thin films [\onlinecite{Exp-FeSe}], LaSrCuO surfaces
[\onlinecite{Bollinger2011}], Li$_x$ZrNCl layered materials [\onlinecite{Exp-LiZrNCl}], as well as graphene-based hybrids [\onlinecite{bouchiat12}];
see also the reviews [\onlinecite{SITReview,Fazio01}]. The experimental studies were complemented by a large body of theoretical work \cite{Fisher90,Fin,Feigelman07,Burmistrov12}. A theory by Fisher and co-authors\cite{Fisher90} describing SIT in terms of vortex condensation and invoking duality between the charge and vortex physics predicted a  single-parameter scaling near the SIT, with a universal resistance $h/4e^2$ at the transition. 
These predictions, are however, in disagreement with many experiments. The actual physics near the SIT is in general more complex and includes also a mutual influence (renormalization) of disorder (that controls the localization effects) and interactions in different channels. The corresponding RG formalism was developed by Finkelstein \cite{Fin} and extended recently in Ref.\onlinecite{Burmistrov12}. 
 
Experimental investigations of SIT in 1D geometry are more scarce. In addition to experiments on JJ chains \cite{ChowEtAl98,HavilandEtAl00,KuoChen01,MiyazakiEtAl02,TakahideEtAl06} mentioned above, the SIT was studied on MoGe nanowires \cite{Bezryadin00,Bollinger08,Rogachev12}. The theory of destruction of superconductivity by QPS in nanowires was developed in Ref.~\onlinecite{Zaikin97}, see also Ref.~\onlinecite{Arutyunov08} for a review. On the qualitative level, the experimental observation of the SIT in 1D structures is consistent with theoretical expectations. On the other hand, attempts to identify the parameter controlling the transition and to characterize the scaling near the SIT in experimental works---both on JJ arrays and on nanowires---have led to contradictory conclusions, largely inconsistent with previous theories. 

Thus,  further work, both theoretical and experimental, is needed in order to understand the physics of SIT in 1D systems, which served as one of motivations for the present paper. On the theory side, one can anticipate, in view of the importance of disorder for the SIT in 2D geometry, that it may play an important role for the physics of SIT in 1D systems as well. 
With these motivations, we investigate in the present work the influence of disorder on the SIT in JJ chains . Our main goals are to determine the SIT phase diagram and to calculate the temperature and length dependence of the conductivity of a disordered JJ chain around the SIT.  While we focus on the JJ chain model, we expect that our results should be to a large extent applicable also to a broader class of 1D systems undergoing the SIT. 

We consider a generic model with a capacitive coupling to the ground ($C_0$) and between the islands ($C_1$), and  study both the cases of short-range ($C_1 \ll C_0$) and long-range  ($C_1\gg C_0$) Coulomb interaction.  Incorporating QPS fluctuations and including the effect of random offset charges, we map the lattice model in the low-energy regime to the sine-Gordon theory. We further take into account a second type of disorder: randomness in the QPS fugacity. Such randomness will generically arise as a result of  interplay of fluctuations in the Josephson energy with random offset charges. Employing the renormalization group (RG), we determine the phase diagram of the system. As may be anticipated on the basis of studies of the persistent current \cite{MatveevEtAl02}, random offset charges weaken the effect of QPS, thus favoring superconducting correlations. On the contrary, phase slips with random fugacity are not weakened by random stray charges and thus widen the regime of insulating behavior. Using the memory-function framework, we calculate the temperature dependence of resistivity $\rho(T)$ (in the long-system limit, $N\to\infty$), as well as the length dependence of the resistance of a finite chain, $R(N)$. In the vicinity of the SIT, both these dependences show pronounced non-monotonic behavior, which reflects the multifaceted physics of the problem at different energy and length scales. 

The paper is structured as follows. In Sec.~\ref{Sec:Model} we introduce the lattice model and present its mapping, in the low-energy sector, to a theory of sine-Gordon type.  The case of short-range interaction  ($C_1 \ll C_0$) is considered in Sec.~\ref{Sec:Short-range}. We first analyze the RG equations (Sec.~\ref{Sub:RG-short-range}) and then study the transport properties (Sec.~\ref{Sub:Transport-short-range}). In Sec.~\ref{Sec:Long-range} we explore the long-range-interaction limit, $C_0 \ll C_1$, relevant for most of experimental realizations of  JJ chains.  Finally, in  Sec.~\ref{Sec:Conclusion} we summarize main results of the paper and compare our findings to available experimental results. 

\section{Model}
\label{Sec:Model}

%%%%%%%%%%%%%%%%%%%%%%%%%%%%
\begin{figure}
\centering
\includegraphics[scale=0.28]{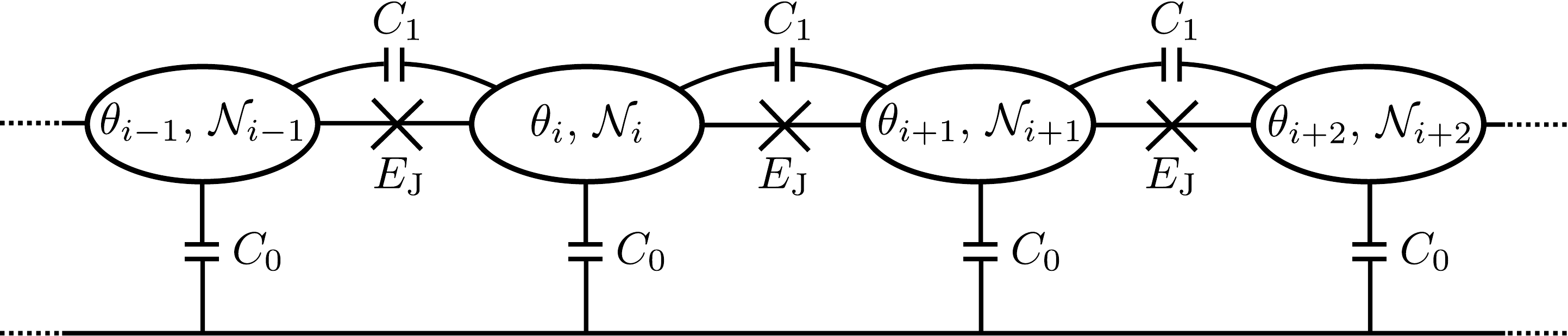}
\caption{Schematic depiction of a Josephson junction chain. The superconducting grains are coupled via tunnel-barriers that provide a capacitance $C_1$ and connected to the ground via capacitances $C_0$. The superconducting phase of an island $i$ is denoted by $\theta_i$, and the corresponding number of Cooper pairs by ${\cal N}_i$.}
\label{Fig:schematics}
\end{figure}
%%%%%%%%%%%%%%%%%%%%%%%%%%%%%%%

We consider a chain of superconducting islands that are smaller than the bulk coherence length so that each of them can be described by a single phase $\theta$.
The system is sketched in Fig.~\ref{Fig:schematics}.
 Weak links between the islands provide Cooper-pair tunneling characterized by the Josephson energy $\Ej$. 
 The Josephson coupling competes  with charging effects described by the  capacitance matrix $C_{ij}$ which we assume  to contain on-site and nearest neighbor capacitances denoted by $C_0$ and $C_1$, respectively. These two capacitances define two charging energy scales $E_0=(2e)^2/C_0$ and $E_1=(2e)^2/C_1$. 
The charging energy $E_0$   plays a key role for the ultimate long-scale behavior of the theory. In particular,  a JJ chain is expected\cite{BradleyDoniach84,ChoiEtAl98} to undergo a quantum superconductor-insulator transition at $K_0\equiv \sqrt{\Ej/E_0}\sim 1$.  On the other hand, the charging energy $E_1$, while irrelevant in the case $C_0\gg C_1$, has strong impact on local properties of the system in the opposite limit, $C_1\gg C_0$. In particular, the parameter $K_1\equiv\sqrt{E_J/E_1}$ 
controls the superconducting correlations   at distances shorter than the screening length of the charge-charge interaction, $\Lambda=\sqrt{C_1/C_0}\gg 1$.   
 
It is convenient to introduce the dimensionless capacitance matrix, $S_{ij}=C_{ij}/C_1$, given by 
\begin{equation}
S_{ij}=\left(2+\frac{1}{\Lambda^2}\right) \delta_{i,j}-\delta_{i,j+1}-\delta_{i,j-1}.
\end{equation}
 The Hamiltonian of a clean JJ chain takes then the form 
 \begin{equation}
 \label{Eq:lattice-model}
\Ham=\frac{E_1}{2}\sum_{i,j}S^{-1}_{ij}{\cal N}_i {\cal N}_j+\Ej \sum_i\left[1-\cos\left(\theta_i-\theta_{i+1}\right)\right],
\end{equation} 
where  ${\cal N}_i$ is the number of Cooper pairs on the $i$-th island canonically conjugate to the superconducting phase,  $[{\cal N}_i,\theta_j]= i  \delta_{i,j}$. 

In this work we focus on low-energy properties of the model \eqref{Eq:lattice-model} and of its generalizations (with disorder included) to be  introduced  below. 
More precisely, we consider modes with momenta $q\lesssim1$ (unless specified explicitly, we measure all distances in units of the lattice spacing) and frequencies $\omega\lesssim \Omega_0$. Here the  frequency cutoff (the width of the plasmonic  band)
\begin{equation}
\Omega_0=\sqrt{\frac{\Ej E_1 E_0}{E_1+E_0}}
\end{equation}
varies from the  frequency $u_0=\sqrt{\Ej E_0}$ of phase oscillations of a single grain to the plasma frequency associated to a single  Josephson junction, 
$\omega_p=\sqrt{\Ej E_1}$,  as the Coulomb interaction range $\Lambda$ changes between zero and  infinity. 

The effective low-energy description of our model  is conveniently formulated in terms of the field $\phi(x)$ related to the Cooper-pair density
by $\pi {\cal N}(x)=- \partial_x\phi(x)$ and, correspondingly,  obeying the commutation relation
\begin{equation}
\left[-\partial_x\phi,\theta(x^{\prime})\right]= i \pi\delta(x-x^{\prime}).
\end{equation}
On the Gaussian level, the (imaginary-time) action for $\phi$ reads
\begin{equation}
S_0=\frac{\Omega_0}{2\pi^2 K}\int\dfrac{\diff q}{2\pi}\dfrac{\diff \w}{2\pi} \left[\frac{\w^2}{\Omega_0^2}+ \frac{\left(1+1/\Lambda^2\right)q^2}{q^2+1/\Lambda^2}\right]\left|\phi(q,\w) \right|^2,
\label{Eq:S0}
\end{equation}
and describes one-dimensional plasma waves with energy dispersion
\begin{equation}
\epsilon(q)=\frac{\wpp |q|}{\sqrt{q^2+1/\Lambda^2}}.
\label{Eq:spectrum}
\end{equation}
Here we have introduced the dimensionless constant 
\begin{equation}
K=\sqrt{\frac{\Ej}{E_0}+\frac{\Ej}{E_1}}=K_1\sqrt{1+\frac{1}{\Lambda^2}}
\end{equation}
that interpolates between $K_0$ for $\Lambda \to 0$ and $K_1$ for $\Lambda\to \infty$. 

In terms of the superconducting phases, the action (\ref{Eq:S0}) describes  small long-wavelength fluctuations of $\theta_i$ around the superconducting ground state, $\theta_i\equiv \theta(x)={\rm const}$,  favored  by the Josephson coupling. It fully captures the physics of the model at low temperatures and in the  semiclassical regime  $\Ej\gg E_0$.  The crucial role in the destruction of the superconducting phase by charging effects is played by quantum phase slips (QPS)---quantum events of $2\pi$ winding of  the phase difference $\theta_{i}-\theta_{i-1}$  on one of the Josephson junctions . 
In the (imaginary-time) path-integral description of the system,  QPS are vortices in the superconducting phase $\theta(x, \tau)$. In order to account for those topological excitations, one needs to add to the quadratic action (\ref{Eq:S0}) a correction
\begin{equation}
S_{\rm{ps}}=\frac{y \Omega_0}{\sqrt{2\pi^3} }\int\diff x\,\diff \tau\, \cos\left[2\phi(x,\tau)\right],
\label{Eq:Sps}
\end{equation}
where $y$  is the dimensionless matrix element for the phase slip (fugacity of a vortex).    Phenomenologically, the correction (\ref{Eq:Sps}) can be understood as follows\cite{GiamarchiBook,BenfattoEtAl12}: the operator $\e^{2  i  \phi(x_0,\tau_0)}$ acts as a translation operator that shifts $\theta(x, \tau)$ after a time $\tau_0$ and for  $x<x_0$ by $2\pi$, creating thus a QPS. A detailed microscopic derivation of Eq.~(\ref{Eq:Sps}) is presented in  Appendix \ref{App:Derivation-theory} for completeness. 

Under the condition $\Ej\gg \min(E_1, E_0)$, superconducting correlations are well developed in the system, at least locally. 
A QPS is then a kind of a tunneling process, and the microscopic  QPS amplitude is exponentially small:
\begin{equation}
y\propto e^{-\alpha K} \,,
\label{Eq:y}
\end{equation}
with a  numerical coefficient $\alpha$ depending on the screening length $\Lambda$. Strictly  speaking, the precise value of $\alpha$ depends also on details of the ultraviolet cutoff that supplements the effective long-wavelength description of the system, Eqs.~(\ref{Eq:S0}) and~(\ref{Eq:Sps}). We refer the reader 
to Refs. \onlinecite{MatveevEtAl02, BradleyDoniach84,ChoiEtAl98,RastelliEtAl13} and Appendix \ref{A:MLG} for the estimates of $\alpha$ in various limiting cases. 
In the rest  of the paper, we treat $y$ as a phenomenological parameter (small in the regime $K\gtrsim 1$) and focus on implications of 
QPS for low-energy properties of the disordered system. 

The main subject of the present paper is the effect of disorder on transport properties of the system. Several sources of disorder in JJ arrays are known. One unavoidable kind of disorder is represented by random stray charges $Q_i$ that ``frustrate'' the charging part of the Hamiltonian:
\begin{equation}
\frac{E_1}{2}\sum_{i,j}S^{-1}_{ij}{\cal N}_i {\cal N}_j \ \rightarrow \  \frac{E_1}{2}\sum_{i,j}S^{-1}_{ij}\left({\cal N}_i-Q_i\right) \left({\cal N}_j-Q_j\right). 
\end{equation}  
The random stray charges influence phase slips via the Aharonov-Casher effect\cite{AharonovCasher}: in the course of a phase slip between island $i$ and $i+1$ the  wave function of the system accumulates the phase factor $\exp( i {\cal Q}_i)$ with
\begin{equation}
{\cal Q}_i=2\pi \sum_{k\leq i}Q_i.
\end{equation}
Correspondingly, in the presence of stray charges the phase-slip action (\ref{Eq:Sps}) transforms into 
\begin{equation}
S_{\rm{ps, Q}}=\frac{y\Omega_0}{\sqrt{2\pi^3}}  \int\diff x\,\diff \tau\, \cos\left[2\phi(x,\tau)-{\cal Q}(x)\right]\,.
\label{Eq:SpsQ}
\end{equation}
Statistical properties of the stray charges $Q$ may depend on various material-dependent aspects. In this work, we assume for simplicity correlations of the stray charges to be short-ranged and describe them by a single parameter, the variance $D_Q$:
\begin{equation}
\langle Q(x) Q(x^\prime)\rangle=\frac{D_Q}{2\pi^2}\delta(x-x^\prime).
\label{Eq:DQ}
\end{equation}

Another source of quenched disorder in a JJ array is fluctuations of the charging and Josephson energies from junction to junction.  These fluctuations can lead to spatial variations of the parameters of the quadratic action (\ref{Eq:S0}). Since such variations do not directly influence the charge transport, we will not take them into account.  In addition, the spatially fluctuating charging and Josephson energies influence locally the value of the QPS amplitude $y$ which is of key importance for transport properties.
Taking into account also the presence of the stray charges (that provide a random phase of the fluctuating term), we model this type of disorder by
\begin{equation}
S_{\xi}=\int\diff x\,\diff \tau\left[\xi(x)\e^{2 i \phi(x,\tau)}+\mathrm{h.c.}\right]
\label{Eq:Sxi}
\end{equation}
with random complex amplitude $\xi$.  In analogy with Eq.~(\ref{Eq:DQ}), we assume that $\xi$ is short-range correlated \cite{note-correlations}, 
\begin{equation}
\left<\xi(x)\xi^{\ast}(x^{\prime})\right>=\frac{u_0^2 D_{\xi}}{(2\pi)^2}\delta(x-x^{\prime}).
\label{Eq:Dxi}
\end{equation}

The complete description of our model reads
\begin{equation}
\label{Eq:Full-action}
S=S_0+S_{\rm{ps, Q}}+S_{\xi},
\end{equation}
where $S_0$, $S_{\rm{ps, Q}}$ and $S_{\xi}$ are given by Eqs. (\ref{Eq:S0}), (\ref{Eq:SpsQ}), and (\ref{Eq:Sxi}), respectively.  
The action (\ref{Eq:Full-action}) constitutes the staritng point for our study of transport properties of disordered JJ chains that we present in Secs. \ref{Sec:Short-range} and 
\ref{Sec:Long-range} for the cases of short-range and long-range  interaction, respectively.  

\section{Short-range Coulomb interaction}
\label{Sec:Short-range}

We first study the system  in the regime $C_1\ll C_0$  when the charge interaction is local in space. In this case the plasma waves have a linear spectrum, and the quadratic part of the action assumes the form of a Luttinger liquid:
\begin{equation}
S_0=\frac{1}{2\pi^2 u_0 K_0}\int \diff x \,\diff \tau \left[ u_0^2\left(\partial_x \phi\right)^2 +  \left(\partial_{\tau} \phi\right)^2\right] \,,
\label{Eq:S0Loc}
\end{equation}
with the Luttinger liquid parameter $K_0=\sqrt{\Ej/E_0}$ and velocity $u_0=\sqrt{\Ej E_0}$. (We remind the reader that, in our notations, distances are measured in units of the lattice spacing, so that dimensions of energy and velocity coincide.) 

We recognize now that the effective description of a disordered JJ chain with local interaction, as provided by Eqs.  (\ref{Eq:Full-action}),   (\ref{Eq:S0Loc}), (\ref{Eq:SpsQ}) and 
(\ref{Eq:Sxi}), is closely related to that of a disordered interacting quantum wire developed  in  Ref. \onlinecite{GiamarchiSchulz88}.  Specifically, the random fugacity term,  Eq. (\ref{Eq:Sxi}), corresponds  to disorder-induced backward scattering in a quantum wire. Further, the uniform  QPS amplitude $y$  can be viewed as describing the effect of a (commensurate) periodic potential on the electronic system. Finally, the stray charges $Q$ play the role of random forward scattering in the quantum-wire problem. 
In what follows, we exploit the similarity between our system and a model of a disordered quantum wire in order to derive the  appropriate RG description. This will allow us to determine the phase diagram of a JJ chain with local charge-charge interaction  and to study the low-temperature transport in the system.

\subsection{RG equations}
\label{Sub:RG-short-range}

In order to derive RG equations for the action (\ref{Eq:Full-action}) with $S_0$ given by Eq.~(\ref{Eq:S0Loc}), we largely follow the approach of Ref.~\onlinecite{GiamarchiSchulz88}.  We use the replica trick to perform the average over the random QPS amplitude $\xi$. On the other hand,  it proves convenient to postpone the average over random stray charges till a later stage of the derivation. Upon the averaging over $\xi$, the action of our replicated theory takes the form
\begin{align}
\label{Eq:S_repl_local}
S&=\sum_{i=1}^{n}\left(S_0[\phi^i]+S_{\rm{ps},Q}[\phi^i]\right)+\sum_{i,j=1}^n S_{\xi}[\phi^i,\phi^j],
\\
S_0[\phi^{i}]&=\frac{1}{2\pi^2 u_0 K_0}\int \diff x\diff \tau \left[u_0^2(\partial_x \phi^i)^2 +(\partial_{\tau}\phi^i)^2\right],
\\
S_{\rm{ps},Q}[\phi^i]&=\frac{y u_0}{\sqrt{2\pi^3}}\!\int \!\!\diff x \diff \tau \cos\left[2\phi^i-2\pi\int_{-\infty}^x \!\!\! \diff  z Q(z)\right],
\label{Eq:S_ps_replicated}
\\
S_{\xi}[\phi^i,\phi^j]\!&=\!-\frac{u_0^2 D_{\xi}}{(2\pi)^2}\!\!\int\!\! \diff x \diff \tau \diff \tau^{\prime}\! \cos\! \left[2\!\left(\phi^i(x,\tau)-\phi^j(x,\tau^{\prime})\right)\! \right]\!.
\end{align}
Here $i = 1,2, \ldots, n$ is the replica index, and the limit $n\to 0$ should be taken.

To construct RG equations, we analyze the correlation function 
\begin{equation}
\label{Eq:correlation_function_R}
R(x_1-x_2,\tau_1-\tau_2)=\left< \e^{ i  2 \phi^{j}(x_1,\tau_1)}\e^{- i  2 \phi^{j}(x_2,\tau_2)}\right>,
\end{equation}
where the angular brackets denote the average with respect to the action \eqref{Eq:S_repl_local} as well as over the the random field $Q(x)$. We calculate this correlation function perturbatively in the phase slip-fugacity $y$ (up to $2^{\rm{nd}}$ order) and in the disorder strength $D_{\xi}$ (up to $1^{\rm{st}}$ order). These perturbative corrections allow us to infer the RG equations, see Appendix \ref{App:RG-short_range} for detail. The result reads:
\begin{eqnarray}
\frac{\diff K_0}{\diff l}&=&-\frac{1}{2}y^2K_0^2 \left[\BesselI_0(D_Q)-\StruveL_0(D_{Q})\right]-\frac{1}{2}K_0^2 D_{\xi},
\label{Eq:RG_K0_local}
\\
\frac{\diff y}{\diff l}&=&(2-\pi K_0)y,
\label{Eq:RG_y_local}
\\
\frac{\diff D_{\xi}}{\diff l}&=&(3-2\pi K_0)D_{\xi},
\label{Eq:RG_Db_local}
\\
\frac{\diff D_{Q}}{\diff l}&=&D_{Q},
\label{Eq:RG_DQ_local}\\
\frac{\diff u_0}{\diff l}&=&-\frac12 u_0 K_0 y^2 r(D_Q)-\frac{1}{2}u_0K_0 D_{\xi},
\label{Eq:RG_u0_local}
\end{eqnarray}
where
\be
r(D_Q) = \StruveL_2(D_Q) - \BesselI_2(D_Q)+\frac{2}{3\pi}D_Q,
\label{Eq:rD}
\ee
$l$ is the logarithm of the running length scale, $\BesselI_n$ denotes the $n$-th modified Bessel function of the first kind, and $\StruveL_n$  is the $n$-th modified Struve function. 
We stress that while Eqs. \eqref{Eq:RG_K0_local} - \eqref{Eq:RG_u0_local} are perturbative in $y$ ($2^{\rm{nd}}$ order) and $D_\xi$ ($1^{\rm{st}}$ order), they are exact in $K_0$ and  $D_{Q}$. 
 
 In the absence of disorder ($D_Q=D_{\xi}=0$), our RG  equations for $K_0$ and $y$ reduce to the standard  Berezinskii-Kosterlitz-Thouless (BKT) form,
 \begin{eqnarray}
 \frac{\diff K_0}{\diff l}&=&-\frac{1}{2}y^2K_0^2,
\label{Eq:RG_K0_localClean}
\\
\frac{\diff y}{\diff l}&=&(2-\pi K_0)y,
\label{Eq:RG_y_localClean}
 \end{eqnarray}
 describing a quantum superconductor-insulator transition at  $\pi K_0^c=2$ for an infinitesimally small fugacity. The flow of the velocity $u_0$ vanishes in the clean limit due to space-time symmetry. 

Let us now analyze the RG flow in a disordered system. If the superconducting correlations  are sufficiently strong, $\pi K_0>2$, the superconducting state is stable with respect to small $y$ and $D_\xi$. For $\pi K_0<2$ phase slips may proliferate in the course of RG.
 The random fugacity perturbation remains irrelevant as long as $\pi K_0> 3/2$, and we first drop it  from our discussion. Although the QPS amplitude $y$ grows under RG for any $\pi K_0<2$,  its impact on the properties of the system depends on the stray charges. 
If the bare charge disorder is sufficiently weak, the coefficient $D_Q$ (growing under RG) remains small at scales where the QPS amplitude becomes of order unity and localization develops. 
On the other hand, examination of Eq.~\eqref{Eq:RG_K0_local} shows that for 
 $D_Q\gg 1$ the correction to $K_0$ induced by QPS is proportional to $y^2/D_Q$. Correspondingly,  in this regime we expect localization effects to proliferate only 
 if $y^2/D_Q$ becomes of order unity. This conclusion receives further support in Sec. \ref{Sub:Transport-short-range}, where we show that exactly the same parameter controls the perturbative corrections to the conductivity of the system.
 
Under the assumption $D_Q\gg1$, the RG  equations simplify to
\begin{eqnarray}
\frac{\diff K_0}{\diff l}&=&-\frac{1}{2}K_0^2 D_{\xi,y},
\label{Eq:RG-short-range-K0-strongDQ}
\\
\frac{\diff D_{\xi,y}}{\diff l}&=&(3-2\pi K_0)D_{\xi,y},
\label{Eq:RG-short-range-Dxiy-strongDQ}
\\
\frac{\diff u_0}{\diff l}&=&-\frac{1}{2}u_0K_0 D_{\xi,y}.
\label{Eq:RG-short-range-u0-strongDQ}
\end{eqnarray}
where $D_{\xi,y}=D_{\xi}+2y^2/\pi D_Q$. 
Equations (\ref{Eq:RG-short-range-K0-strongDQ})--(\ref{Eq:RG-short-range-u0-strongDQ}) correspond to results of Giamarchi and Schulz\cite{GiamarchiSchulz88} for the case of a 1D system of spinless particles with backward-scattering disorder, with identification of our $\pi K_0$ to $K$ of Ref.~\onlinecite{GiamarchiSchulz88}.
We see that strong random stray charges effectively make the ``regular'' QPS contribution 
(\ref{Eq:SpsQ}) indistinguishable from the random-fugacity one, Eq. (\ref{Eq:Sxi}).  
In particular, although formally derived under the condition $y\ll 1$,  the RG equations (\ref{Eq:RG_K0_local}), (\ref{Eq:RG_y_local}),  (\ref{Eq:RG_DQ_local}), and (\ref{Eq:RG_u0_local}) remain valid at $D_Q\gg 1$ in a much wider range $y^2/D_Q\ll 1$.
The critical value of $K_0$ where the superconductor-insulator transition takes place (at vanishing $y$) is changed by stray charges from $2/\pi$ to $3/2\pi$. 
The reduction of the effect of QPS on the properties of the system under strong forward scattering has a simple physical interpretation: in the presence of stray charges $Q$ the QPS do not add up coherently since they experience destructive interference. 

\begin{figure}
\centering
\includegraphics[scale=0.65]{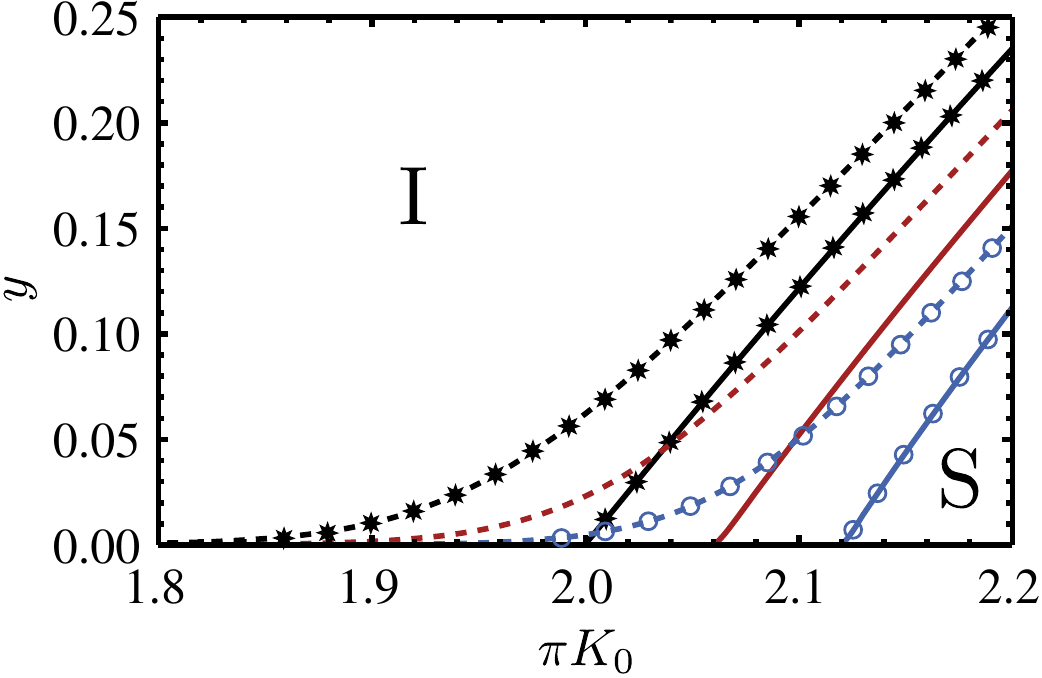}
\caption{Phase diagram for a JJ chain with short-range Coulomb interaction in the $\pi K_0$ -- $y$ plane. The chain is in the insulating phase (I) to the left of each transition line,  and in the superconducting phase (S) to the right of it. The black solid line (with stars) corresponds to the clean case ($D_{\xi}=D_{Q}=0$). The other two solid curves describe a chain without random stray charges ($D_Q=0$) but with random QPS fugacity: $D_{\xi}=0.1$ (red), $D_{\xi}=0.2$ (blue, open circles). The dashed curves include a small amount of stray charges ($D_Q=10^{-12}$), with the same value of $D_{\xi}$ as on the solid line with the same color (same symbol).}
\label{Fig:PhaseDiagr_local}
\end{figure}

 To establish the phase diagram of the model, we solve RG equations \eqref{Eq:RG_K0_local}-\eqref{Eq:RG_DQ_local} numerically.  
We work in the plane spanned by $\pi K_0$ and the QPS amplitude $y$, treating them as independent parameters [see a discussion around  Eq.~(\ref{Eq:y})].  The resulting phase diagram is shown in Fig.~\ref{Fig:PhaseDiagr_local}.
The area to the left of each transition line corresponds to the parameter regime where the system is insulating, while the regime to the right corresponds to the superconducting phase at zero temperature. The black solid line (with stars) ending at $\pi K_0=2$ separates superconducting and insulating  phases in a clean system. The other two solid lines---red (no symbol) and blue (open circles)---illustrate the shift of the transition due to a non-zero (but relatively small) randomness $D_\xi$ in the QPS amplitude.  As expected, increasing this kind of disorder shifts the transition point to the right, i.e., in favor of the insulating regime. On the contrary, random stray charges have the opposite effect. The dashed lines in Fig. \ref{Fig:PhaseDiagr_local} show the phase boundary in the presence of a small amount of stray charges. It is seen that even a very small value of the stray-charge disorder $D_Q$ shifts quite appreciably the phase boundary, enhancing the superconductivity.

%%%%%%%%%%%%%%%%%%%%%%%%%%%%%%%
%%%%%%%%%%%%%%%%%%%%%%%%%%%%%%%

\subsection{Transport}
\label{Sub:Transport-short-range}

We are now in a position to study the low-temperature transport properties of our model (in the case of short-range interaction).
The current-operator can be deduced from the continuity equation $\partial_t \rho_e+\partial_x j_e=0$, where $\rho_e=-(2e/\pi) \partial_x \phi$. We therefore find
\begin{equation}
j_e=\frac{2e}{\pi}\partial_t \phi=2eu_0K_0 \partial_x \theta.
\end{equation}
Using the Kubo formalism, the conductivity can be expressed as
\begin{equation}
\label{sigma-w}
\sigma(\w)=\frac{ i }{\w}\left[4e^2  u_0   K_0+\chi(\w)\right],
\end{equation}
where 
\begin{equation}
\chi(\w)=-\int \diff x \int_{-\infty}^{t} \!\!\!\diff t^{\prime}\e^{ i  \w (t-t^{\prime})}\left<\left[j_e(x,t),j_e(x^{\prime},t^{\prime}) \right] \right>
\end{equation}
is the retarded current-current correlation function. In the clean limit and in the absence of phase slips, the DC conductivity is infinite:
\begin{equation}
\sigma(\w)=4\pi e^2 u_0 K_0 \left[\delta(\w)+\frac{ i }{\pi}\mathcal{P}\frac{1}{\w} \right].
\end{equation}
Both types of phase-slip processes (homogeneous and random fugacity) yield a finite DC limit. 
To compute the DC conductivity, we use the memory function formalism \cite{GoetzeWoelfle72,Giamarchi91,RoschAndrei00,MirlinPolyakovVinokur07}. A finite conductivity in the zero frequency limit implies, according to Eq.~(\ref{sigma-w}),  $\chi(0)=-4e^2u_0 K_0$.  By introducing the meromorphic memory function
\begin{equation}
\label{Eq:memory-function}
M(\w)=\frac{\w\chi(\w)}{\chi(0)-\chi(\w)},
\end{equation}
the conductivity can be expressed as
\begin{equation}
\label{Eq:cond-memory}
\sigma(\w)= i  4e^2u_0 K_0\frac{1}{\w+M(\w)}.
\end{equation}
It is further convenient to introduce the correlation function
\begin{equation}
C(\w) =\int \diff x\int_0^{\infty}\diff t\, \e^{ i  \w t}\left<\left[F(x,t),F(0,0)\right] \right> \,,
\label{Cw}
\end{equation}
where
\begin{equation}
F(x,t) =\left[ \Ham,j_e(x,t) \right]\,.
\end{equation}
The average in Eq.~(\ref{Cw}) can be performed at $y=0$ and $D_{\xi}=0$ if one is only interested in the lowest order of perturbation theory. The memory function can now be expressed as
\begin{equation}
\label{Eq:memory_with_C}
M(\w)=\frac{1}{-\chi(0)}\frac{C(\w)-C(0)}{\w}.
\end{equation}

Perturbative computation of the memory function to the lowest non-trivial order in $y$ and $D_\xi$ can be carried out in full analogy with Ref.~\onlinecite{Giamarchi91,MirlinPolyakovVinokur07}.  Details of this calculation are presented for completeness in  Appendix \ref{App:memory_function}. To obtain the conductivity, we combined the RG procedure, which allows us to renormalize the theory up to the infrared cutoff set by the temperature or by the system size, with the perturbative evaluation of the memory function.

\subsubsection{Clean limit}
\label{Subsub:Clean}

%%%%%%%%%%%%%%%%%%%%%%%%%%%%%%%%%%%%%%%%%%%%%%%%%%%%%%%%%%%%%%%%%
\begin{figure}
\centering
\includegraphics[scale=0.7]{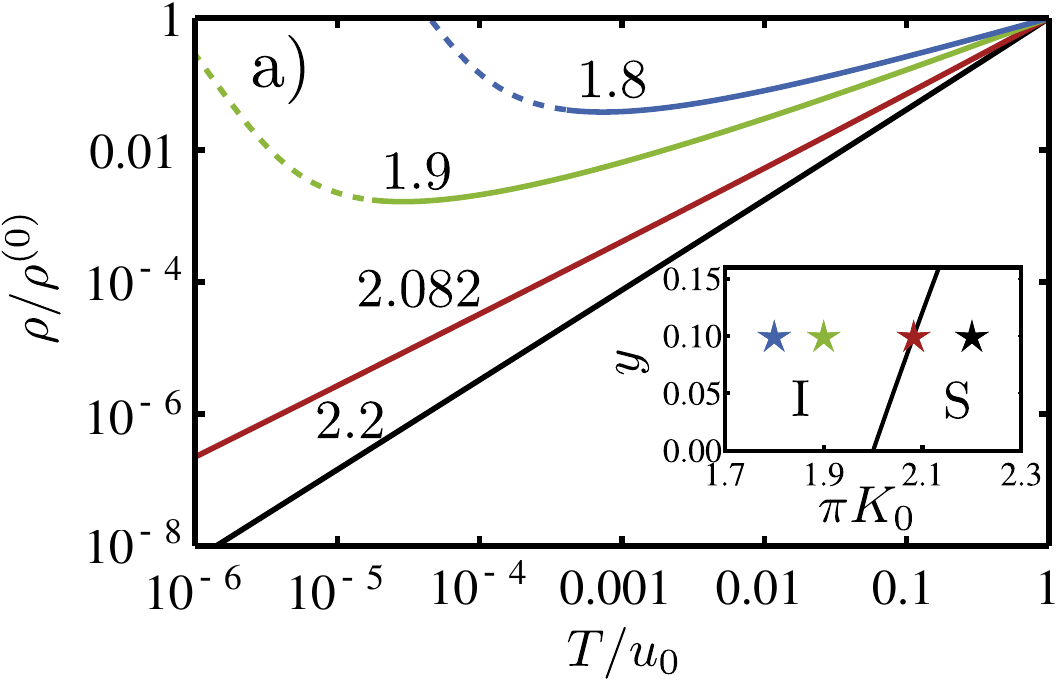}

\vspace{4mm}

\includegraphics[scale=0.7]{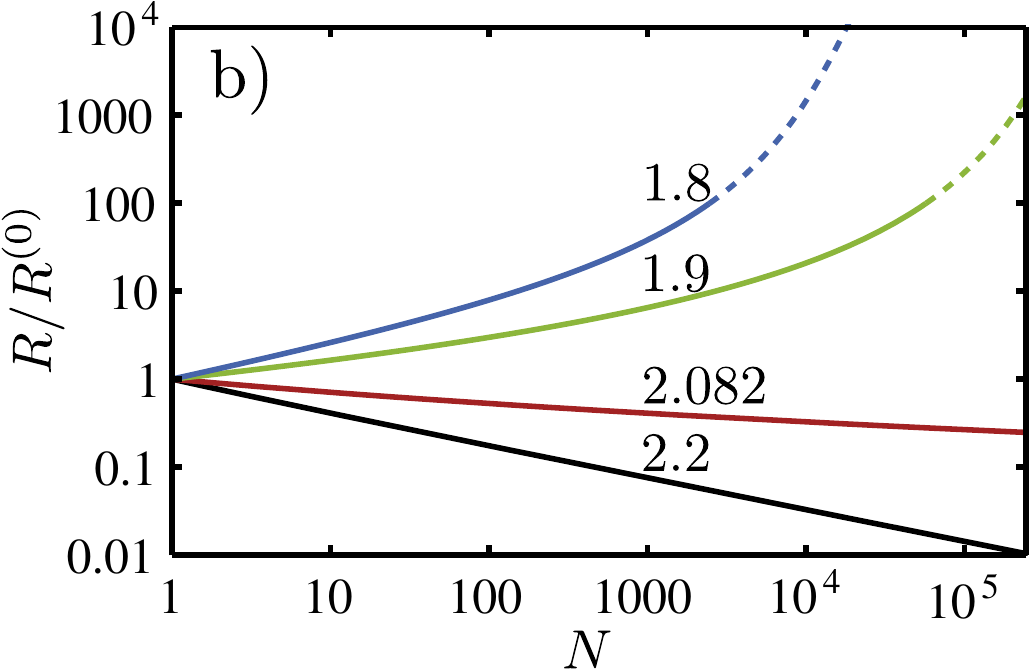}
\caption{a) Temperature-dependent resistivity in the clean case for short-range Coulomb interaction. The numbers on the curves indicate the value of $\pi K_0$. The value of the fugacity is the same for every curve, $y=0.1$. Inset: Phase diagram in the $\pi K_0$-$y$-plane. The stars mark the position of the corresponding resistivity curve with the same color. To the left of the black line, the chain is in the insulating phase (I), while to the right it shows superconducting correlations (S). The dashed parts are qualitative extrapolations illustrating the flow towards the insulating (infinite resistivity) fixed point. b) Dependence of the resistance with array length $N$ at $T=0$ with the same parameters   as for the resistivity plot.}
\label{Fig:CleanCondShortRange}
\end{figure}
%%%%%%%%%%%%%%%%%%%%%%%%%%%%%%%%%%%%%%%%%%%%%%%%%%%%%%%%%%%%%%%%%%%%

We start the analysis of the DC conductivity with the clean limit, $D_{\xi}=D_Q=0$. Our model, Eqs. (\ref{Eq:S0Loc}) and (\ref{Eq:Sps}), can be 
 viewed in that case as describing interacting fermions in a periodic potential. The physical process behind the finite resistance  is then the Umklapp scattering.  It is known \cite{RoschAndrei00}  that under {\it incommensurate} filling the Umklapp  scattering processes induced by interaction and periodic potential are extremely inefficient and lead to exponentially large conductivity in the system. This is not the issue in the present case, however, as our periodic potential is commensurate. On the perturbative level the resulting conductivity reads in the static limit:
\begin{equation}
\sigma(T)=\frac{8e^2a}{y^2 h}\frac{\Gamma^2(\pi K_0)}{\Gamma^4(\pi K_0/2)}\left(\frac{2\pi a T}{u_0}\right)^{3-2\pi K_0},
\label{Eq:CleanCond}
\end{equation}
where we have restored explicitly the  lattice spacing $a$. Incorporating renormalization effects transforms the power-law temperature dependence of conductivity, Eq.~(\ref{Eq:CleanCond}) into a more complex behavior. To establish it, we renormalize the theory from the original ultraviolet cutoff $a$ down to the thermal length $N_{\rm th}(T)=u_0/T$ where the RG, Eqs.~ (\ref{Eq:RG_K0_localClean}) and (\ref{Eq:RG_y_localClean}), is terminated. Since the velocity $u_0$ itself gets renormalized, this implies the following equation for the corresponding RG scale $l^{\ast}(T)$:
\be
\e^{l^{\ast}}=\frac{u_0(l^{\ast})}{T}.
\ee
Combining this renormalization with Eq.~(\ref{Eq:CleanCond}) yields the following behavior of the conductivity with temperature~$T$:
\begin{equation}
\sigma(T) \sim \frac{u_0[l^{\ast}(T)]}{T y^2[l^{\ast}(T)]},
\label{Eq:CleanCondScaling}
\end{equation}
 The symbol ``$\sim$'' in Eq.~(\ref{Eq:CleanCondScaling}) and in analogous formulas below means ``up to a numerical coefficient of order unity''.
In the clean case the velocity $u_0$ is not renormalized, so that $l^{\ast}(T)=\ln\left(u_0/T\right)$. It is convenient to normalize the conductivity by its bare value as $\sigma^{(0)}=\sigma(T=u_0)$. The temperature dependence of the correspondingly normalized resistivity $\rho / \rho^{(0)}$, with $\rho  =1/\sigma$ and $\rho^{(0)} =1/\sigma^{(0)}$,
is shown for the clean case in Fig.~\ref{Fig:CleanCondShortRange}a.  
If we are in the  superconducting regime (black curve, $\pi K_0=2.2$), the resistivity decreases witch decreasing temperature. In the insulating regime (green and blue curve, $\pi K_0=1.9$ and $1.8$) the resistivity shows a strongly non-monotonic dependence. Specifically, $\rho(T)$ decreases at relatively high temperatures (quite similarly to superconducting curves) because the growth of $y^2$ needs to overcome the additional factor $1/T$ in Eq.~\eqref{Eq:CleanCondScaling}. Therefore, the resistivity starts to increase only at lower temperatures, where $K_0$ is renormalized below $3/2\pi$. Since our treatment is perturbative in $y$, we have to stop the renormalization when $y\sim 1$. Since the sine-Gordon theory in the clean case can be mapped onto a fermionic system with umklapp scattering, we expect an RG flow towards the Mott-insulator fixed point (infinite resistivity) if umklapp scattering is relevant. We therefore plot as a dashed curve the extrapolation extracted from the RG beyond the perturbative regime in order to show the qualitative tendency at low temperatures. The parameters of the red curve ($\pi K_0=2.082$) lie on the transition line. Using the BKT equations, we analytically find
\begin{equation}
\label{Eq:Crit-resistivity-short-range}
\rho^{\mathrm{crit}}(T)/\rho^{(0)}=\frac{T/u_0}{\left[1+(\pi K_0 -2)\ln(u_0/T)\right]^2}
\end{equation}
for the resistivity on the critical line, where $K_0$ is assumed to be not to far from $K_0^c=2/\pi$. 

The resistivity $\rho(T)$ calculated above characterizes the problem in the thermodynamic limit, $N \to \infty$.  This corresponds to the situation when the system size $N$ is much larger than the thermal length $N_{\rm th}(T)$, so that the infrared cutoff for the renormalization effects is provided by the temperature $T$, while the dependence of the resistance on $N$ is simply Ohmic. It is important to analyze also the opposite situation, $N \ll N_{\rm th}(T)$. The appropriate characteristics of the system in this case is the length-dependent resistance $R(N)$ at zero temperature. 
To determine it, we renormalize the theory until the cutoff reaches the system length $N$ and then make use of the relation $R=\rho \cdot N$. The result reads
\begin{equation}
R(N) \sim \frac{h}{e^2}y^2[l=\ln N].
\end{equation}
The resulting dependences $R(N)$ are presented in Fig.~\ref{Fig:CleanCondShortRange}b. All curves are normalized by the resistance in the ultraviolet: $R^{(0)}=R(N=1)$. The parameters (bare values of $y$ and $K_0$) are identical to the resistivity curves of Fig.~\ref{Fig:CleanCondShortRange}a. Curves in the insulating regime show an increasing resistance with system size (green and blue, $\pi K_0=1.9$ and $1.8$), while for the superconducting (black, $\pi K_0=2.2$) and the critical (red, $\pi K_0=2.082$) curves the resistance decreases. The critical curve has the following system-size dependence  [cf. Eq.~\eqref{Eq:Crit-resistivity-short-range}]:
\begin{equation}
\label{Eq:Rcrit}
R^{\mathrm{crit}}(N)/R^{(0)}=\frac{1}{\left[1+(\pi K_0-2)\ln(N)\right]^2},
\end{equation}  
i.e., the resistance at criticality drops with increasing $N$ in a logarithmically slow fashion. This implies that curves that are on the insulating side but very close to the critical line (not shown in the figure) will show a non-monotonic dependence: $R(N)$ will first decrease with increasing $N$ and only then will start increasing. This non-monotonicity in ``weakly insulating'' dependences $R(N)$ is, however, much less pronounced than that in the corresponding $\rho(T)$ curves; the difference is related to the additional factor of $T$ in the $T$-dependence in Eq.~(\ref{Eq:Crit-resistivity-short-range}) in comparison with the $N$-dependence in Eq.~(\ref{Eq:Rcrit}). 

\subsubsection{Disordered system: Random stray charges}
\label{Subsub:Stray}

After having analyzed the clean limit, we now discuss the effect of random stray charges. Using the results from  Appendix \ref{App:memory_function}, we obtain on the level of the perturbation theory
\begin{equation}
\sigma(T) \sim 
\begin{cases}
\displaystyle  \frac{e^2 a}{h y^2}\left(\frac{2\pi a T}{u_0}\right)^{3-2\pi K_0}, & D_Q u_0 /aT \ll 1,
\\[0.5cm]
\displaystyle e^2  a\frac{D_Q}{h y^2}\left(\frac{2\pi a T}{u_0}\right)^{2-2\pi K_0}, & D_Q u_0 /aT \gg 1.
\end{cases}
\label{Eq:CondpsShort-range}
\end{equation}
Already at this stage, we see that the power of $T$ is reduced by unity for the case of strong random stray charges, as compared to the regime in which stray charges are weak. Incorporating renormalization effects, we find the temperature dependence of conductivity:
\begin{equation}
\frac{\sigma(T)}{\sigma^{(0)}}\sim
\begin{cases}
\displaystyle \frac{y_0^2}{y^2(T)}\frac{u_0(T)}{T}, &D_Q(T) \ll 1,
\\[0.5cm]
\displaystyle \frac{y_0^2 D_Q(T)}{y^2(T)}\frac{u_0(T)}{T}, & D_Q(T) \gg 1 \,,
\end{cases}
\label{Eq:ScalingCondDQ}
\end{equation}
where $y_0$ is the bare value of the fugacity. Since the velocity is renormalized when random stray charges are present, we need to solve the equation $\e^{l^{\ast}}=u_0(l^{\ast})/T$ numerically to find $l^{\ast}(T)$. 

%%%%%%%%%%%%%%%%%%%%%%%%%%%%%
\begin{figure}
\centering
\includegraphics[scale=0.7]{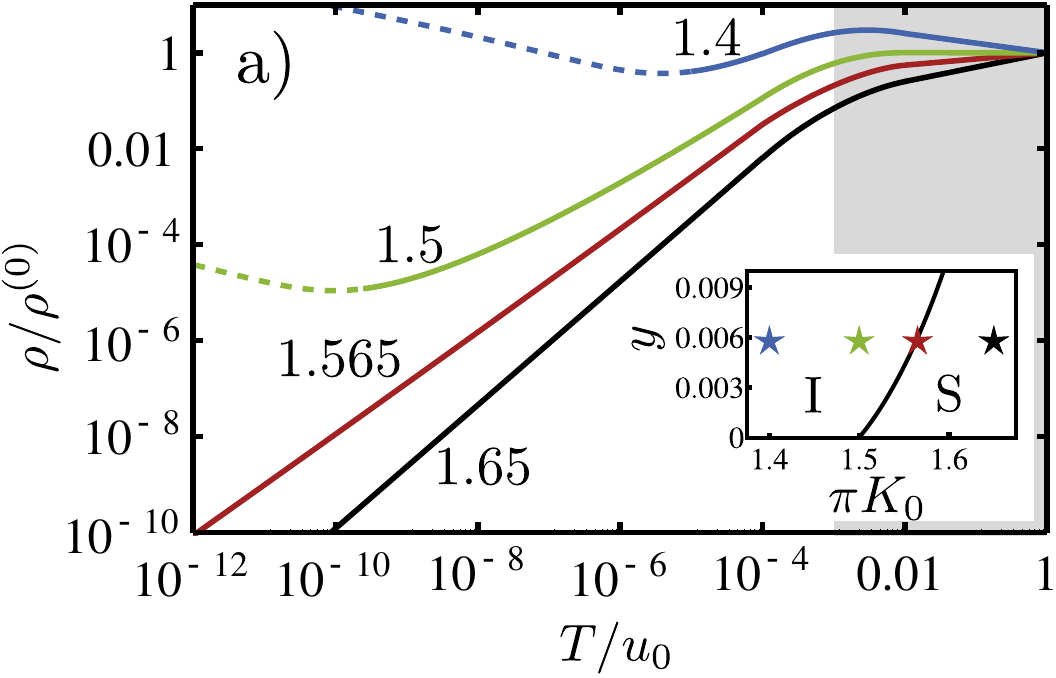}

\vspace{4mm}

\includegraphics[scale=0.7]{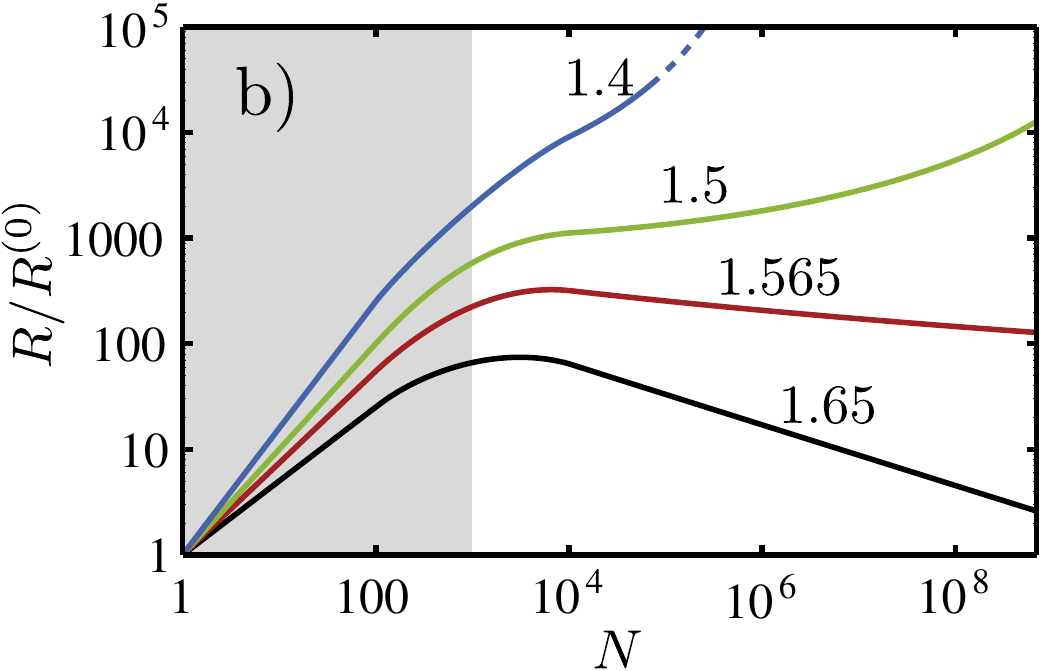}
\caption{a) Scaling of the resistivity with temperature in the presence of random offset charges for short-range Coulomb interaction. The numbers at the curves indicate the values of $\pi K_0$. The other parameters $D_Q=10^{-3}$ and $y=6\cdot 10^{-3}$ are the same for all curves. In the gray region random stray charges are weak (the boundary is at the temperature $T_Q$). The black curve ($\pi K_0=1.65$) corresponds to a superconducting system (although phase slips are relevant). The green and blue curve ($\pi K_0=1.5$ and $1.4$) are in the insulating regime. The dashed blue and green lines, which correspond to the temperature range $T<T_{\mathrm{ps}}$, represent extrapolations to demonstrate the tendency at lowest temperatures (flow towards the Anderson insulator). Inset: Phase diagram in the $\pi K_0$-$y$-plane. The stars show the position of the parameters for the resistivity plots in the phase diagram. To the left of the black line, the system shows insulating behavior (I), while to the right it shows superconducting correlations (S). b) Length-dependent resistance at $T=0$ for the same parameters as for the resistivity curves. The characteristic scales, which correspond to the temperature scales $T_Q$ and $T_{\mathrm{ps}}$ in the panel a), are $N_Q$ (boundary of gray region) and $N_{\mathrm{ps}}$ (beginning of of the dashed line, seen on the blue curve only).}
\label{Fig:WeakDQCondShort-Range}
\end{figure}
%%%%%%%%%%%%%%%%%%%%%%%%%%%

The scaling behavior of the resistivity $\rho=1/\sigma$ including random offset charges is depicted in Fig.~\ref{Fig:WeakDQCondShort-Range}a.
For high temperatures (gray region) the curves are similar to the clean case,  Fig.~\ref{Fig:CleanCondShortRange}, for the same values of $K_0$.
(It is worth emphasizing that the values of $K_0$ used in Fig.~\ref{Fig:WeakDQCondShort-Range} are smaller than those in Fig.~\ref{Fig:CleanCondShortRange} since stray charges shift the SIT phase boundary, as is seen from the comparison of the insets of both figures.)

 There is a crossover temperature, $T_Q/u_0=D_Q^{(0)}$, where the renormalized strength of stray charges [$D_Q(l)$] reaches a value of order unity. If  the system is still in the perturbative regime at this temperature ($y \ll 1$), the resistivity is suppressed at lower temperatures by an additional small factor $1/D_Q(T) \propto T$ according to the second line of Eq.~\eqref{Eq:ScalingCondDQ}. In the plot we used an interpolation formula to match smoothly the two regimes [$D_Q(T)$ smaller and larger than unity, or, equivalently, $T$ above and below $T_Q$] of Eq.~\eqref{Eq:ScalingCondDQ}. 

As Eq.~\eqref{Eq:CondpsShort-range} indicates, the perturbative parameter in the strong disordered case, $D_Q(T)\gg 1$, is $y^2(T)/D_{Q}(T)$ rather than $y^2(T)$. This confirms the conclusion made in Sec.~\ref{Sub:RG-short-range} that random stray charges stabilize superconducting correlations.
Thus, in the regime of strong stray charges there is a competition between them and the phase slips. 
On the superconducting side of the SIT, which is represented  by  the black curve ($\pi K_0=1.65$), the stray charges are sufficiently strong to suppress the effect of phase slips. On the other hand, on the insulating side represented by the green ($\pi K_0=1.5$) and blue ($\pi K_0=1.4$) curves, phase slips win over stray charges. In this case the resistivity shows an upturn at low temperatures. We expect that for lower temperatures (beyond the perturbative regime) the resistivity will continue to grow because the proliferation of phase slips will destroy the superconducting correlations and the quantum localization will take over. To illustrate qualitatively this behavior, we perform an extrapolation of the RG into the strong-coupling regime (shown by dashed lines). 

As is clear from the above discussion, the temperature dependence in the insulating regime is strongly non-monotonic and in general consists of three regions of temperatures with alternating signs of $d\rho/dT$. Such a behavior is well pronounced for the blue curve ($\pi K_0=1.4$). At high temperatures $T \gtrsim T_Q$ (gray region), the resistivity grows with lowering temperature. In this regime the disorder is weak and the strong growth of the phase-slip fugacity results in $d\rho/dT < 0$. Below the temperature $T_Q$, random stray charges suppress the influence of phase slips, which leads to a decrease of resistivity with lowering temperature, $d\rho/dT > 0$ . However, phase slips grow strongly enough to overcome the suppression by stray charges. The corresponding upturn of $\rho(T)$ is visible only in the strong-coupling regime (shown by dashed lines), since we need to renormalize down to $K_0=1/\pi$ to overcome the additional power of $T$ [cf. ~Eq.~\eqref{Eq:CondpsShort-range}]. This happens at a temperature $T_{\mathrm{ps}}$ which can be estimated as
\begin{equation}
T_{\mathrm{ps}}\sim u_0 \left(y^{(0)}\right)^{\frac{2}{3-2\pi K_0}}\left(\frac{u_0}{T_Q}\right)^{\frac{1}{3-2\pi K_0}}
\end{equation}
assuming $\pi K_0$ is not too close to $3/2$. The temperature range $T<T_{\mathrm{ps}}$ is shown on the blue and green curves in Fig.~\ref{Fig:WeakDQCondShort-Range}a by dashed line. For $y^{(0)} \ll 1$ the two scales $T_Q$ and $T_{\mathrm{ps}}$ are distinct, $T_{\mathrm{ps}}\ll T_Q$.

The red curve, $\pi K_0=1.565$, is the SIT phase boundary.  We determine the
temperature dependence of resistivity at this critical line for sufficiently low temperatures, $T \lesssim T_Q$, where the stray charges are important. Solving the corresponding RG equations Eqs.~\eqref{Eq:RG-short-range-K0-strongDQ}-\eqref{Eq:RG-short-range-u0-strongDQ}, we find (the renormalization of the velocity can be neglected close to the critical point $K_0^c=3/2\pi$):
\begin{equation}
\rho^{\mathrm{crit}}(T)\sim \frac{T/u_0}{[1+(\pi K_0^Q-3/2)\ln(T_Q/T)]^2},
\label{rho-crit}
\end{equation}
where $K_0^Q=K_0[l=\ln N_Q]$ is the renormalized value of $K_0$ at the mean free path $N_Q=1/D_Q^{(0)}$.

To determine the dependence of the resistance $R(N)$ on the system size for $N < N_{\rm th}(T)$, we terminate the renormalization by $N$, which yields 
\begin{equation}
\frac{R(N)}{R^{(0)}}\sim
\begin{cases}
\displaystyle \frac{y^2[\ln(N)]}{y_0^2}, &N\ll 1/D_Q^{(0)},
\\[0.5cm]
\displaystyle \frac{y^2[\ln(N)]}{y_0^2 D_{Q}^{(0)} N}, & N \gg 1/D_Q^{(0)}.
\end{cases}
\label{Eq:salingResistanceShortRanged}
\end{equation}
The dependences $R(N)$  are shown in Fig.~\ref{Fig:WeakDQCondShort-Range}b where the same values of the parameters as in Fig.~\ref{Fig:WeakDQCondShort-Range}a are used. In analogy with the $\rho(T)$ plot, we interpolate in the intermediate regime where $D_Q \sim 1$ to obtain a smooth matching of the two limits of Eq.~\eqref{Eq:salingResistanceShortRanged}.
For the chosen parameters, the insulating curves [green ($\pi K_0=1.5$) and blue ($\pi K_0=1.4$)] show a monotonically increasing behavior. For chains that are longer than the mean free path $N_Q$, the growth is weakened  (in an intermediate range of $N$) since phase slips can no longer interfere coherently. At the scale $N_{\mathrm{ps}} \sim u_0/T_{\mathrm{ps}}$, which is the correlation length at which the system enters the strong coupling regime, the resistance growth is accelerated again. The superconducting curve (black, $\pi K_0=1.65$) shows a strongly non-monotonic behavior of $R(N)$.
Specifically,
for small chain sizes (shorter than mean free path, gray region) the resistance is increasing, so that one could think that the system is in the insulating phase. However, for larger systems, the resistance starts to decrease:  at $T=0$ and in the thermodynamic limit the parameters of the black curve correspond to the superconducting regime (see inset of Fig.~\ref{Fig:WeakDQCondShort-Range}a). The red curve ($\pi K_0=1.565$) is on the transition line and shows qualitatively the same behavior as the black curve ($\pi K_0=1.65$). The decrease at large $N$ is, however, much weaker. We find for chain sizes larger than the mean free path $N_Q$
\begin{equation}
\label{R-crit}
R^{\mathrm{crit}}(N)\sim \frac{1}{[1+(\pi K_0^Q-3/2)\ln(N/N_Q)]^2}
\end{equation} 
for the length dependence of the resistance on the phase boundary.

\subsubsection{Disordered system: Random stray charges and random fugacity}
\label{Subsub:RandomFug}
 
We are now going to analyze how disorder that produces phase slips with random fugacity influences the transport characteristics. It is clear that this kind of disorder reduces the conductivity in contrast to random stray charges. The total memory function is now the sum of the contributions from phase slips with homogeneous and random fugacity, see Appendix \ref{App:memory_function}. The conductivity of the system thus reads
\begin{equation}
\sigma =\frac{1}{\sigma_{\rm ps}^{-1}+\sigma_{\xi}^{-1}},
\end{equation}
where $\sigma_{\rm ps}$ originates from homogeneous phase slips and can be calculated using Eq.~\eqref{Eq:CondpsShort-range}; $\sigma_{\xi}$ originates from phase slips with random fugacity (Appendix \ref{App:memory_function}),
\begin{equation}
\label{sigma-xi}
\sigma_{\xi}(T) \sim \frac{e^2 a}{h D_{\xi}}\left(\frac{2\pi a T}{u_0}\right)^{2-2\pi K_0}.
\end{equation}
Using $\sigma_0=e^2/h$ as a reference conductivity  and performing the renormalization from the bare ultraviolet cutoff $a$ to the thermal length $N_{\rm th}(T)$, we find
\begin{equation}
\frac{\sigma_{\xi}(T)}{\sigma_0}\sim \frac{u_0(T)}{T D_{\xi}(T)}
\end{equation}
and
\begin{equation}
\frac{\sigma_{\rm ps}(T)}{\sigma_0}\sim
\begin{cases}
\displaystyle \frac{1}{ y^2(T)}\frac{u_0(T)}{T}, & D_Q(T) \ll 1,
\\[0.5cm]
\displaystyle \frac{D_Q(T)}{y^2(T)}\frac{u_0(T)}{T}, & D_Q(T) \gg 1.
\end{cases}
\end{equation}

%%%%%%%%%%%%%%%%%%%%%%%%%%%%
\begin{figure}
\centering
\includegraphics[scale=0.7]{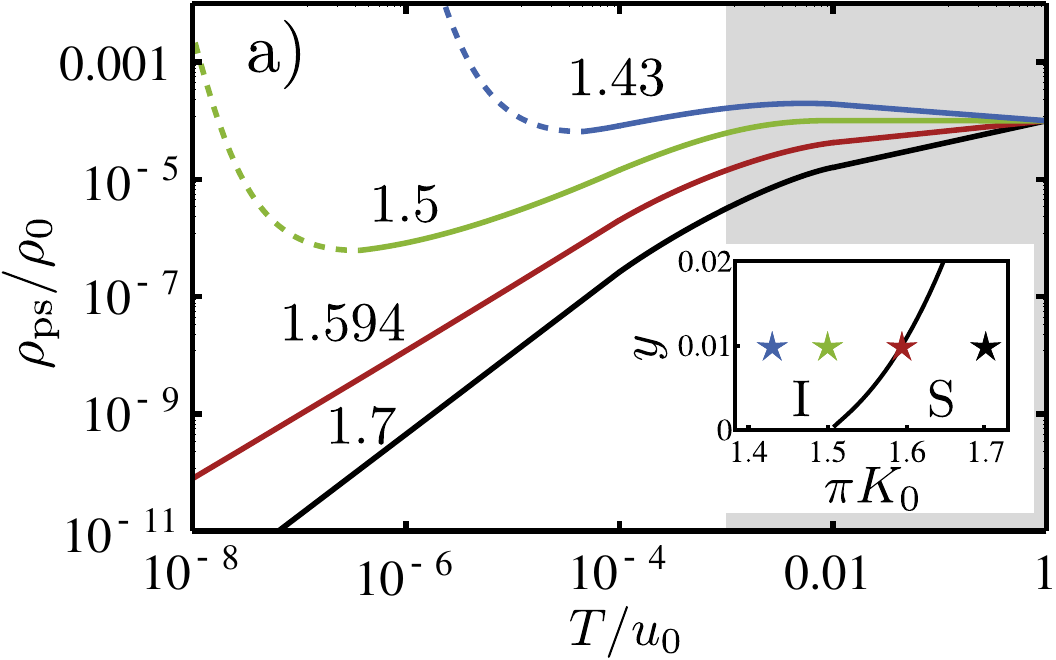}

\vspace{4mm}

\includegraphics[scale=0.7]{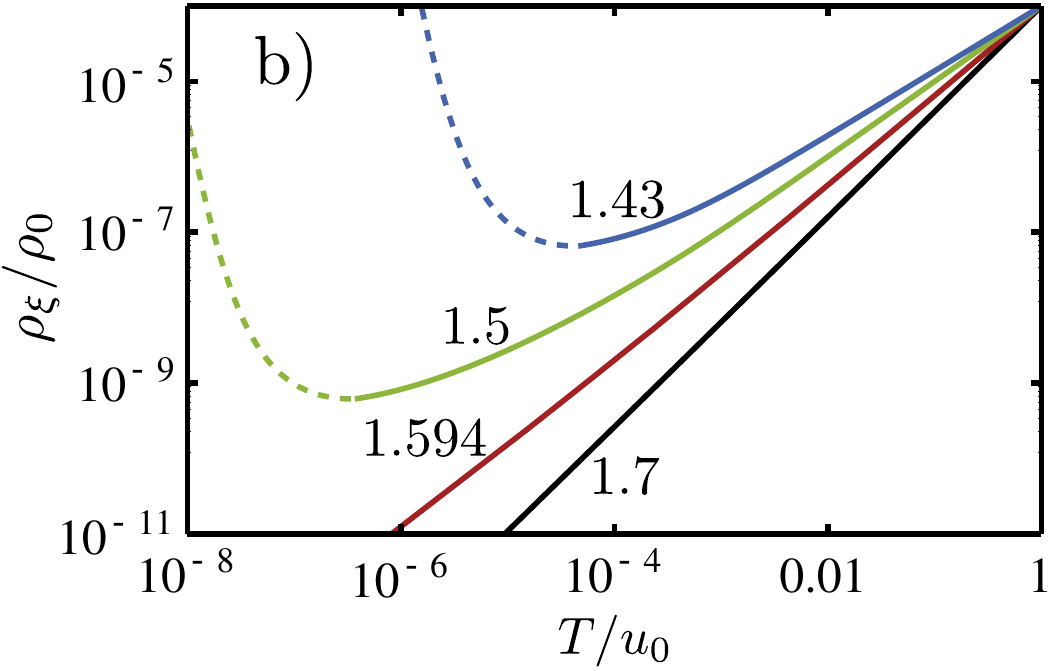}
\caption{Scaling of the temperature-dependent resistivity contributions originating from phase slips with homogeneous fugacity (a) and from phase slips with random fugacity (b). The numbers at the curves indicate the value of $\pi K_0$. The other parameters are the same for all curves: $y=10^{-2}, D_Q=10^{-3}$ and $D_{\xi}=10^{-4}$. The dashed lines at low temperatures are extrapolations illustrating the flow towards the insulating (infinite resistivity) fixed point.  The inset in a) shows the position of the systems in the phase diagram. Parameters that lie to the right of the black line show superconducting correlations (S), while those to the left are characterized by insulating behavior (I).}
\label{Fig:condWholeSysShort-Range} 
\end{figure}
%%%%%%%%%%%%%%%%%%%%%%%%%%%%%%

The contributions to the resistivity from phase slips with homogeneous and random fugacity, $\rho_{\rm ps}=1/\sigma_{\rm ps}$ and $\rho_{\xi}=1/\sigma_{\xi}$, are shown in Fig.~\ref{Fig:condWholeSysShort-Range}. The total resistivity of the system is given by the sum of both contributions, $\rho=\rho_{\mathrm{ps}}+\rho_{\xi}$. In each of the panels, the black curve ($\pi K_0=1.7$) corresponds to the superconducting phase, the green ($\pi K_0=1.5$) and blue ($\pi K_0=1.43$) curves  to the insulating phase, and the red curve ($\pi K_0=1.594$) to the SIT phase boundary; see the phase diagram in the inset of Fig.~\ref{Fig:condWholeSysShort-Range}a.   For the insulating phase, both contributions $\rho_{\rm ps}$ and $\rho_{\xi}$ show a non-monotonic behavior. In the case of random phase slips, the disorder $D_\xi$, which corresponds to  impurity induced backscattering processes in the context of 1D fermionic systems\cite{GiamarchiSchulz88}, does not grow fast enough at the first stage of RG, yielding a  decreasing $\rho_{\xi}$. At a lower temperature, this behavior is reverted, and the system starts to flow towards the localization fixed point (as shown by dashed lines in the figure). The temperature dependence of the $\rho_{\rm ps}$ contribution, which typically dominates the total conductivity, is still more complex and is similar to the case of the only stray-charge disorder, Fig.~\ref{Fig:WeakDQCondShort-Range}a. 
(This similarity is not so surprising, since the random QPS term is effectively generated at large scales by  stray charges and regular QPS.) 
In particular, the blue curve ($\pi K_0=1.43$) in Fig.~\ref{Fig:condWholeSysShort-Range}a exhibits three regions with alternating signs of $d\rho/dT$. Specifically, the resistivity $\rho_{\rm ps}$ incresases with lowering $T$ above the temperature $T_Q$ defined by $D_Q(T_Q) \sim 1$. Below $T_Q$, one first observes a decrease of $\rho_{\rm ps}$, since random stray charges suppress the effect of quantum phase slips. However, with further lowering the temperature, phase slips take over, and the resistivity starts again to increase.

As pointed out above, the contribution $\rho_{\mathrm{ps}}$ is usually larger than $\rho_{\xi}$. This is because (i) the fugacity fluctuations on the UV scale are expected not to exceed the average fugacity and (ii) this inequality is further enhanced by renormalization, see Eqs.~\eqref{Eq:RG_y_local} and \eqref{Eq:RG_Db_local}. In principle, one can imagine a model with a bare value of $y^2$ much smaller than that of $D_{\xi}$, in which case $\rho_{\xi}$ would become important. It remains to be seen whether such a model may be physically relevant in the context of JJ chains.

\section{Long-range Coulomb interaction}
\label{Sec:Long-range}

In the previous Sections, we have presented a detailed analysis of the transport properties of a disordered JJ chain with short-range Coulomb interaction, $\Lambda\ll 1$.  We now turn to the analysis of the model (\ref{Eq:Full-action})  in the opposite limit $\Lambda\gg 1$ relevant to many experimental realizations of the system. As discussed in Sec. \ref{Sec:Short-range},  the random QPS term (\ref{Eq:Sxi}) is effectively generated at large scales by an interplay of  stray charges and regular QPS.  In view of this and for simplicity of the presentation, we assume in this Section that  the bare magnitude of the random QPS term is zero, $D_\xi=0$. 

For the purpose of the RG analysis, it is convenient to recast the action  \eqref{Eq:S0} into a different form (see Appendix \ref{App:Derivation-theory}):
\bea
S_0 &=& \frac{1}{2\pi^2 K}\!\int_{-1}^1\!\frac{\diff q}{2\pi} \int_{-\Omega_0}^{\Omega_0} \frac{\diff \w}{2\pi} \nonumber
\\
& \times & \left[\frac{\w^2}{\Omega_0}+\frac{\Omega_0 q^2}{q^2(1-\ug)+\ug}\right]\!\left|\phi(q,\w)\right|^2.
\label{Eq:S0Final}
\eea
The parameter $\ug=1/(1+\Lambda^2)$ has the meaning of the group velocity of the plasmons (measured in units of $\Omega_0$) at the cutoff momentum $q=1$, while $K$ is given by
\begin{equation}
K=\sqrt{\frac{\Ej(E_1+E_0)}{E_1E_0}}.
\label{Eq:KDef}
\end{equation}
We will see below that $K$ plays the role of the effective Luttinger-liquid parameter (the phase stiffness) at the cutoff.  

\subsection{RG treatment}
\label{Sub:RG-long}

We start our analysis with the presentation  of the RG equations valid for arbitrary screening  length $\Lambda$. 
We sketch here only the main points of the derivation and refer the reader to Appendix \ref{App:RGLong} for details.

An elementary step of our RG consists in the (perturbative-in-$y$) elimination of the modes $\phi(\omega, q)$ with
\begin{equation}
1-dl<q< 1 \qquad  {\rm or} \qquad (1-\ug dl)\Omega_0 <\omega<\Omega_0 \,,
\end{equation}  
with the subsequent rescaling of momentum and energy to restore the initial cutoffs. 
The peculiarity of the present case is that the Gaussian action (\ref{Eq:S0Final}) contains irrelevant perturbations and  its parameters $K$ and $\ug$ are renormalized even to the zeroth order in the fugacity. 
Specifically, we get the following RG equations:
\begin{eqnarray}
\frac{\diff K}{\diff l}=-K\left(1-\ug\right),
\label{Eq:KLongZero}
\\
\frac{\diff \ug}{\diff l}=2 \ug\left(1-\ug\right).
\label{Eq:ugLongZero}
\end{eqnarray}
Equations (\ref{Eq:KLongZero}) and (\ref{Eq:ugLongZero})  have a line of (stable) fixed points with $K={\rm const}$ and  $\ug=1$ describing a generic JJ chain with finite $\Lambda$ in the infrared limit and an (unstable) fixed point $K=\ug=0$  corresponding to a system with infinite-range Coulomb interaction. 

The scaling of the QPS amplitude $y$  imposed by the Gaussian action (\ref{Eq:S0Final}) is given by (see Appendix \ref{App:BareScaling})
\begin{equation}
\frac{\diff y}{\diff l}=\frac{1+\ug}{2} \left(2-\pi K\right)y.
\label{Eq:yLongRangeScaling}
\end{equation}
The factor $(1+\ug)$ in Eq. (\ref{Eq:yLongRangeScaling}) reflects  the engineering dimension of $y$, while the factor  $\left(2-\pi K\right)$ shows that the parameter $K$ can be interpreted as the phase stiffness at the cutoff. 

Equations (\ref{Eq:KLongZero}), (\ref{Eq:ugLongZero}) and (\ref{Eq:yLongRangeScaling}) summarize the scaling properties of the parameters $K$, $\ug$ and $y$ to the lowest order of perturbation theory. 
In Appendix \ref{App:LongRangeSecondOrder} we extend the perturbative treatment of the model to the second order in $y$ and show that in the presence of random stray charges  the resulting RG equations take the form
\begin{align}
\frac{\diff K}{\diff l}&=-(1-\ug)K-\frac{1}{2}y^2 K^2 (1+\ug) \frac{\BesselI_1(D_Q)-\StruveL_1(D_Q)}{D_Q},
\label{Eq:KLong}
\\
\begin{split}
\frac{\diff \ug}{\diff l}&=2\ug(1-\ug)+\frac{y^2}{2}K(1+\ug)\ug\\
&\hspace{-0.8cm}\times \!\!\left[\!(1+\ug)\frac{\BesselI_1(D_Q)-\StruveL_1(D_Q)}{D_Q}-\ug \left[\BesselI_0(D_Q)-\StruveL_0(D_Q)\right]\!\right]\!.
\end{split}
\label{Eq:ugLong}
\end{align} 
Equations (\ref{Eq:yLongRangeScaling}), (\ref{Eq:KLong}), and (\ref{Eq:ugLong}) constitute the main result of this subsection. We will use them in Sec. \ref{Sec:transportLong} to study  the low-temperature transport properties of the system.

It is easy to see that, according to Eq. (\ref{Eq:ugLong}),   $1-u_g\propto y^2 K$ in the infrared limit. Thus, within our accuracy, the RG equations reduce to
\begin{align}
\frac{\diff y}{\diff l}&=\left(2-\pi K\right)y,
\label{Eq:yLongRangeScalingApprox}\\
\frac{\diff K}{\diff l}&=-(1-\ug)K-y^2 K^2  \frac{\BesselI_1(D_Q)-\StruveL_1(D_Q)}{D_Q},
\label{Eq:KLongApprox}\\
\begin{split}
\frac{\diff \ug}{\diff l}&=2(1-\ug)+y^2K\\
&\times\left[2\frac{\BesselI_1(D_Q)-\StruveL_1(D_Q)}{D_Q}- \BesselI_0(D_Q)+\StruveL_0(D_Q)\right].
\end{split}
\label{Eq:ugLongApprox}
\end{align}

It is worth emphasizing that the equations (\ref{Eq:yLongRangeScaling}) and (\ref{Eq:KLong}) are equivalent to Eqs. (\ref{Eq:RG_y_local}) and  (\ref{Eq:RG_K0_local})
(for $D_\xi=0$) under the identification $K_0=K\sqrt{\ug}$.  We also recover Eq. (\ref{Eq:RG_u0_local}) from Eq. \eqref{Eq:ugLongApprox} setting $u_0=\Omega_0/\sqrt{u_g}$ 
apart from the additional term $-(1-u_0^2/\Omega_0)$ on the right hand side arising due to a slightly different renormalization scheme employed in the present Section. 
Thus, Eqs. (\ref{Eq:yLongRangeScalingApprox}), (\ref{Eq:KLongApprox}) and (\ref{Eq:ugLongApprox}) automatically
 capture the correct long-distance physics studied in Sec.~\ref{Sec:Short-range}.
At the same time they predict new features in the behavior of the system at intermediate length-scales $1\ll N \ll K$ where, according to Eq.~\eqref{Eq:yLongRangeScaling}, the phase-slip amplitude experiences a fast drop discussed previously in Ref. \onlinecite{ChoiEtAl98}. 
 In the next subsection (Sec.~\ref{Sec:transportLong}) we will discuss implications of these phenomena for the low-temperature transport properties of the system. 
 
\subsection{Transport in a JJ array with long-range interaction}
\label{Sec:transportLong}

%%%%%%%%%%%%%%%%%%%%%%%%%
\begin{figure}
\includegraphics[scale=0.7]{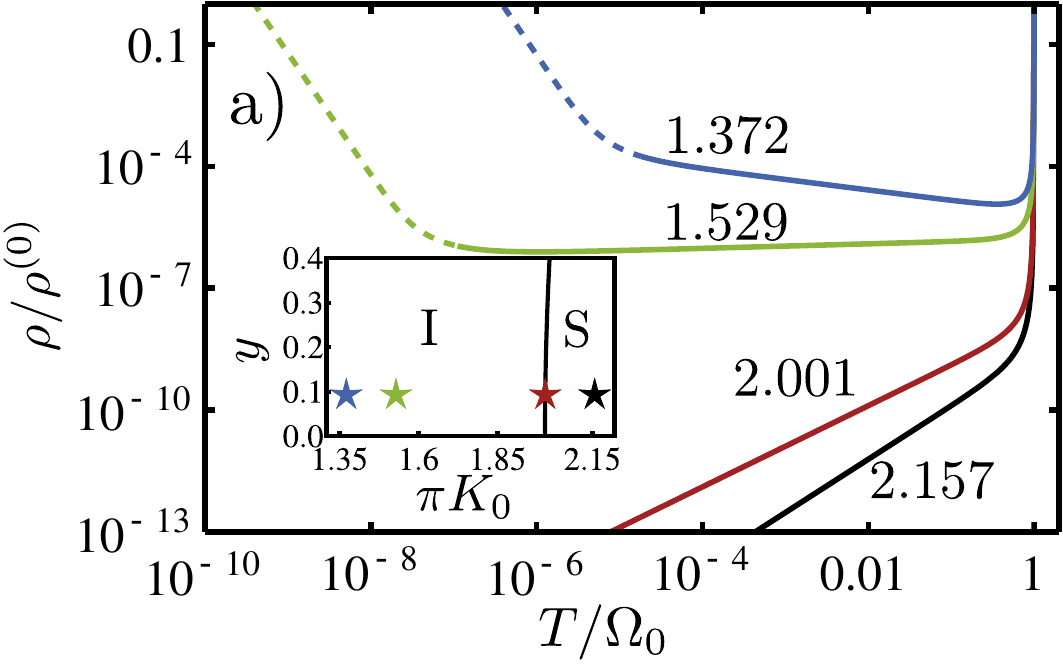}

\vspace{4mm}

\includegraphics[scale=0.7]{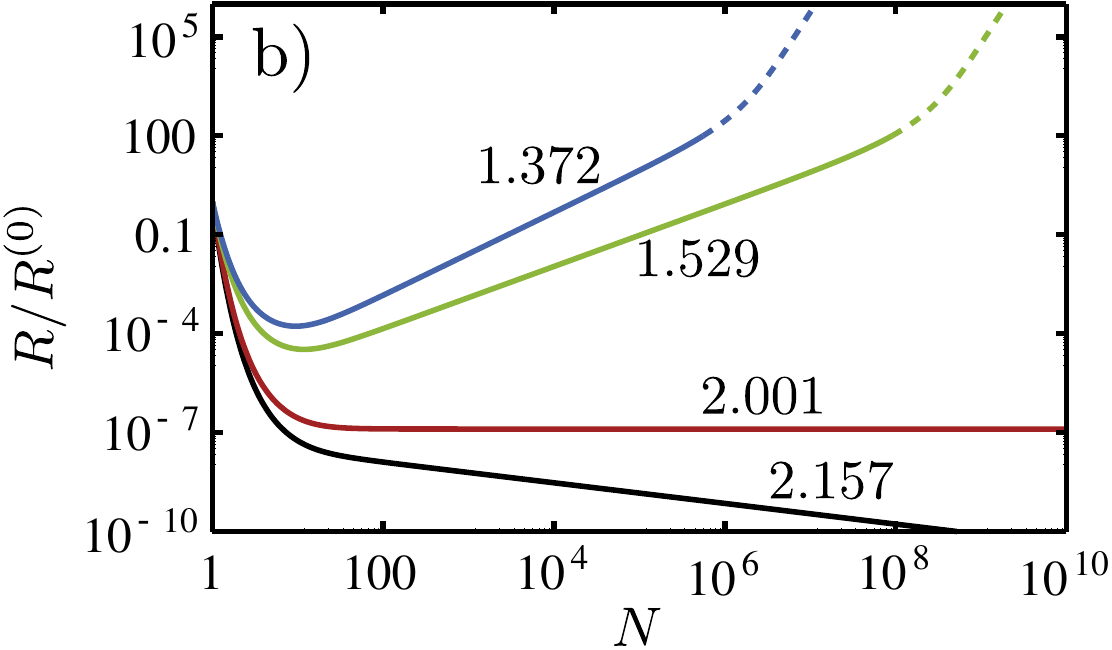}
\caption{Temperature dependence of the resistivity (a) and length dependence of the zero-temperature resistance (b) of a clean JJ chain with screening length $\Lambda=10$ and ultraviolet 
QPS amplitude $y=0.1$. The numbers at the curves indicate the value of $\pi K_0$ controlling the infrared scaling of the QPS amplitude. The dashed lines at low temperatures are extrapolations illustrating the flow towards the insulating (infinite resistivity) fixed point.
The inset shows the position of each of the curves in the phase diagram of the system. 
}
\label{RhoCleanLambda10}
\end{figure}
%%%%%%%%%%%%%%%%%%%%%%%%%

%%%%%%%%%%%%%%%%%%%%%%%%%
\begin{figure}
\includegraphics[scale=0.7]{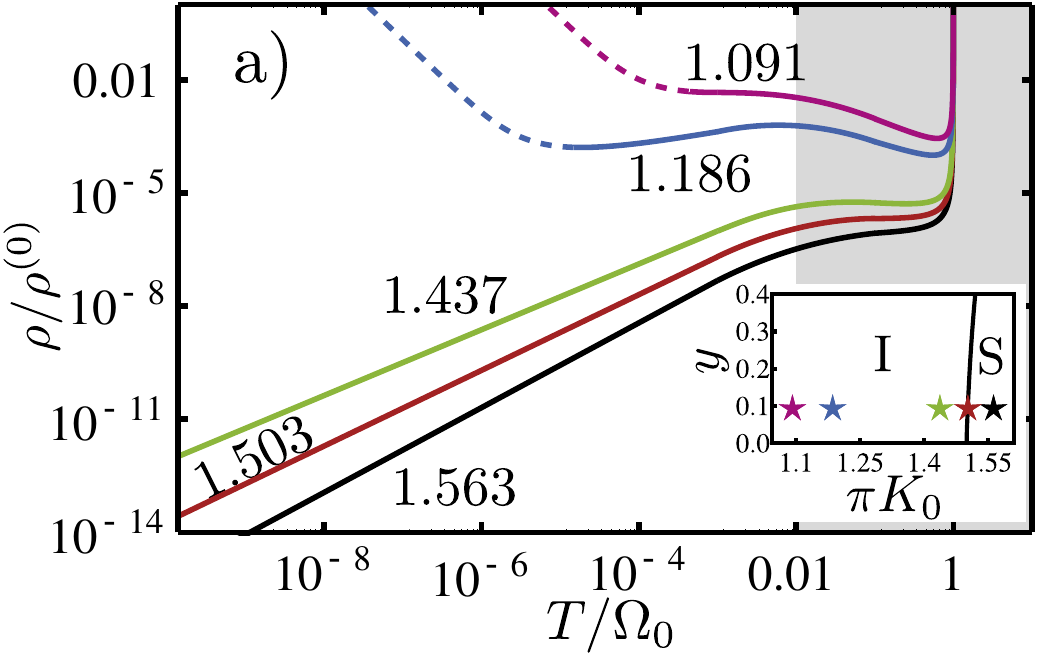}

\vspace{4mm}

\includegraphics[scale=0.7]{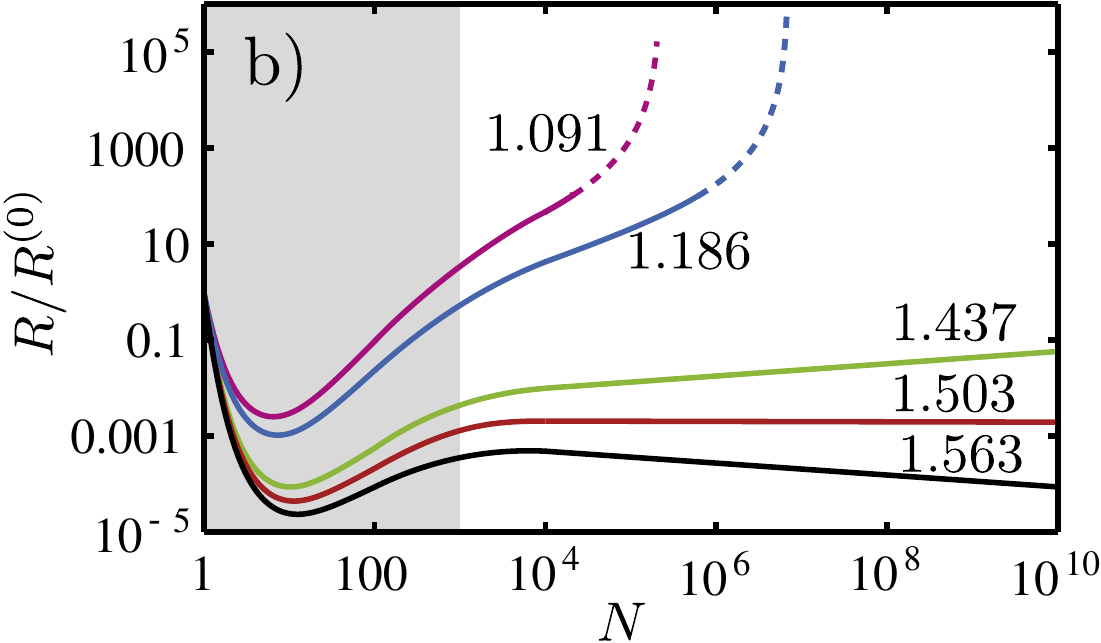}
\caption{Temperature dependence of the resistivity (a) and length dependence of the zero-temperature resistance (b) of a  JJ chain with screening length $\Lambda=10$,  weak random stray charges $D_Q=10^{-3} $, and ultraviolet 
QPS amplitude $y=0.1$. The numbers at the curves indicate the value of $\pi K_0$ controlling the infrared scaling of the QPS amplitude. The dashed lines at low temperatures are extrapolations illustrating the flow towards the insulating (infinite resistivity) fixed point. The inset shows the position of each of the curves in the phase diagram of the system. }
\label{RhoDisTransLambda10}
\end{figure}
%%%%%%%%%%%%%%%%%%%%%%%%%%

In analogy with the case of short-range interaction, Sec. \ref{Sub:Transport-short-range}, we supplement now the RG equations (\ref{Eq:yLongRangeScaling}), (\ref{Eq:KLong}), and (\ref{Eq:ugLong})  
by the expressions for the conductivity of the system  obtained from the memory-function formalism,
 \begin{equation}
\frac{\sigma(T)}{\sigma^{(0)}}\sim
\begin{cases}
\displaystyle \frac{y_0^2 N_{\rm th}(T) }{y^2(T)}, &D_Q(T) \ll 1,
\\[0.5cm]
\displaystyle \frac{y_0^2D_Q(T)N_{\rm th}(T)}{y^2(T)}, & D_Q(T) \gg 1 \,,
\end{cases}
\label{Eq:ScalingCondDQLong-range}
\end{equation}
where $N_{\rm th}(T)$ is the thermal length at which the RG flow is stopped by finite temperature. In the intermediate regime $D_Q(T) \sim 1$ we interpolate between both limits.
The scaling equation for temperature reads (we remind the reader that our RG scheme preserves the energy cutoff $\Omega_0$)
\begin{equation}
\frac{d T (l)}{dl}=u_g T (l).
\label{Eq:T-rescaling}
\end{equation}
The length $N_{\rm th}(T)$ is then the scale where the renormalized temperature $T(l)$ reaches the cutoff $\Omega_0$.  To avoid confusion, we emphasize that all our results yield the resistivity $\rho(T)$ as a function of the physical temperature $T$, which yields the starting point for the RG flow $T(l)$, i.e., $T(l=0) = T$. 
 
The RG equations (\ref{Eq:yLongRangeScaling}), (\ref{Eq:KLong}) and (\ref{Eq:ugLong}) predict, in general, a strongly non-monotonic 
temperature dependence of resistivity and length dependence of the resistance of the JJ chain.
Figure  \ref{RhoCleanLambda10}a  shows the temperature dependence of  a clean ($D_Q=0$) JJ chain for fixed $\Lambda=10$ at various values of $K_0$.  The corresponding scaling of the zero-temperature resistance is shown in Fig.~\ref{RhoCleanLambda10}b.  
Both in the superconducting and insulating phases, the  resistivity shows a rapid drop at temperatures of the order of the cutoff $\Omega_0$ due to the proliferation of local superconducting correlations. Upon lowering the temperature, the system enters into the local regime where the scaling of the QPS amplitude is governed by the infrared stiffness $K_0$. In the vicinity of the critical point $\pi K_0=2$ the resistivity of the system first continues to drop not only in the superconducting phase phase  (black curve, $\pi K_0=2.157$, in Fig.  \ref{RhoCleanLambda10}a) but also at criticality (red line, $\pi K_0=2.001$) and in an adjacent part of the localized phase (green curve, $\pi K_0=1.529$) due to the factor $N_{\rm th }(T)$ in Eq.~\eqref{Eq:ScalingCondDQLong-range}. On the critical line (red curve, $\pi K_0=2.001$) the low-temperature scaling of the resistivity is given by Eq.~\eqref{Eq:Crit-resistivity-short-range}. 
In the insulating phase (green, $\pi K_0=1.529$, and blue, $\pi K_0=1.372$, curves), the localization develops at lowest temperatures. 
 The scaling of the zero-temperature resistance with the system size, Fig.~\ref{RhoCleanLambda10}b, offers a more direct visualization of the superconductor-insulator transition, since the critical curve is characterized by an almost constant resistance [see Eq.~ \eqref{Eq:Rcrit}].
 
Finally, we incorporate the effects of random stray charges, which makes the temperature and system-size dependence of the resistivity even more intricate, see Fig.~\ref{RhoDisTransLambda10}. At short scales (gray region) the effect of the stray charges is negligible and the scaling of the transport characteristics of the system is similar to the clean case. At lower temperatures  or larger system sizes, the interplay of stray charges and phase slips leads to non-monotonic dependences $\rho(T)$ and $R(N)$ with three different regions of behavior, in analogy with  the case of short-range interaction, see Secs.~\ref{Subsub:Stray}, \ref{Subsub:RandomFug} and Figs.~\ref{Fig:WeakDQCondShort-Range}, \ref{Fig:condWholeSysShort-Range}. In total, the curves may show as much as four different regions of behavior, so that the overall dependences $\rho(T)$ and $R(N)$ are in general quite involved and strongly non-monotonic. These distinct regions are clearly seen in Fig.~\ref{RhoDisTransLambda10}. Specifically, at short length scales $N$ (or relatively high temperatures) the resistivity or resistance drop quickly due to the scale dependence of the Luttinger-liquid parameter $K$. For larger scales the phase slips start to play a role and enhance the resistance. At still larger $N$ the effect of phase slips gets suppressed by stray charges. Finally, for the insulator side of the SIT [magenta ($\pi K_0=1.091$), blue ($\pi K_0=1.186$), and green ($\pi K_0=1.437$) curves in Fig. \ref{RhoDisTransLambda10})],  the phase slips blow up at longest scales, driving the system into the insulating fixed point. This complex, strongly non-monotonic behavior makes an experimental identification of the SIT on the basis of experimental data (available for a limited range of $T$ and $N$) a highly non-trivial task. We will compare our results with available experimental data in Sec.~\ref{Sec:Conclusion}.

\section{Summary and discussion}
\label{Sec:Conclusion}

To summarize, we have studied the transport around the SIT in disordered JJ chains. We have started from a lattice model that describes a chain of superconducting islands with a capacitive coupling to the ground ($C_0$) as well as between the islands ($C_1$) and mapped it onto a theory of the sine-Gordon (disordered-Luttinger-liquid) type. This low-energy theory includes QPS fluctuations as well as two types of disorder: random stray charges and randomness in the QPS fugacity.  We have considered both limits of short-range ($C_1 \ll C_0$) and long-range  ($C_1\gg C_0$) Coulomb interaction and studied the resistance of the system by using the RG approach. 

The fixed point of the SIT is of the BKT type and is characterized by the Luttinger-liquid constant $\pi K_0 = 3/2$ and by zero effective fugacity, $D_{\xi,y} =0$ (strength of random phase slips) that controls the resistivity in the presence of offset charges.  
It is worth emphasizing that even a tiny amount of stray charges shifts essentially the SIT boundary in favor of the superconducting phase, see Fig.~\ref{Fig:PhaseDiagr_local}.  The fact that the disorder promotes the superconductivity may seem counterintuitive at first sight. It is interesting to mention that an enhancement of superconductivity by random potential was also found in 3D and 2D systems \cite{Feigelman07,Burmistrov12}, although the mechanism in the present case is different. 

At the critical line (separating the superconducting and insulating phases in the RG flow diagram) the resistivity of an infinite system vanishes linearly with temperature (with a logarithmic correction), Eq.~(\ref{rho-crit}), while the zero-temperature resistance approaches zero logarithmically with increasing system length $N$, Eq.~(\ref{R-crit}). The overall dependences $\rho(T)$ and $R(N)$ are, however, considerably more complex and show several distinct regimes. Specifically, for the case of a short-range interaction, curves belonging to the insulating phase exhibit in general three regimes of behavior taking place consecutively with lowering $T$ (or increasing $N$), see Figs.~\ref{Fig:condWholeSysShort-Range}  and  \ref{Fig:WeakDQCondShort-Range}. At relatively high $T$ (or small $N$) the QPS fluctuations lead to increase of resistivity. At lower $T$ random stray charges become important and suppress the effect of QPS. However, with further lowering $T$, the Luttinger-liquid constant $\pi K_0$ gets renormalized below the critical value 3/2, so that random QPS become relevant, driving the system into the insulating fixed point. For the superconducting phase, the first two of these regimes show up. In the case of a long-range interaction, an additional high-temperature regime emerges, where $\rho(T)$ and $R(N)$ quickly drop with lowering $T$ (respectively, increasing $N$), Fig.~\ref{RhoDisTransLambda10}.

The curves $\rho(T)$ and $R(N)$ around the SIT have thus strongly non-monotonic character: the $T$ and $N$ dependences in the intermediate regimes is essentially different from the ultimate low-$T$ (large-$N$) asymptotics. This makes the experimental determination of the transition point a rather difficult task. Indeed, experimental data are usually obtained in a quite restricted range of $N$ and temperatures, so that the observed behavior may still differ strongly from the infrared asymptotics. Below we briefly discuss the existing experimental data and their interpretation provided in experimental papers, and compare them with our findings. 

The most detailed experimental investigation of the SIT in JJ chains was carried out in Ref.~\onlinecite{ChowEtAl98} where the resistance of arrays (made of Al, with Al$_2$O$_3$ tunnel barriers)  with a length $N$ up to the maximal value $N=255$ was studied in the temperature range from 1\:K down to 50\:mK. The junctions had a SQUID geometry, and the SIT was tuned by the magnetic field. The array was designed in such a way that the screening length $\Lambda$ was quite large, $\Lambda \simeq 10$. The obtained set of $R(T)$ curves for an array with a maximal length ($N=255$), Fig. 3 of Ref.~\onlinecite{ChowEtAl98} is quite similar to our theoretical results (see, in particular, Figs.~\ref{Fig:WeakDQCondShort-Range}a and \ref{RhoDisTransLambda10}a of the present work). Experimental curves that are well on the insulating side show a non-monotonic dependence (first increase with lowering $T$, then decrease, and then again increase), in similarity with our findings. It is tempting to identify the positions of the maximum and minimum on these $R(T)$ curves, $400$ mK and $100$ mK, as corresponding to $T_Q$ and $T_{\mathrm{ps}}$, respectively. An independent determination of the bare values of $y$ and $D_Q$ would be needed to verify this identification. We also note that the low-$T$ minimum is not observed on insulating $R(T)$ curves for shorter chains, $N=63$, which implies that they are way too short to probe the large-$N$ behavior.  Relatively short system sizes may also explain why the quantitative criterion for the SIT deduced in Ref.~\onlinecite{ChowEtAl98} does not conform to the theory. Specifically, the authors of Ref.~\onlinecite{ChowEtAl98} have concluded that their experimental data imply an SIT at (in our notations) $\pi K_0 = 2 /\sqrt{\Lambda}$ (i.e., at $\pi K_0 \simeq 2/3$ for their value of $\Lambda$). This is in disagreement with our theory that yields a transition at $\pi K_0 = 3/2$. We speculate that chain lengths $N$ in the experiment were probably not sufficiently large and/or the temperature was not low enough to probe the actual SIT. In other words, the results in Ref.~\onlinecite{ChowEtAl98} were likely substantially affected by intermediate regimes analyzed in our work. 
%In this connection, it is worth mentioning that the crossing point of curves corresponding to arrays with different $N$, Fig. 4 of Ref.~\cite{ChowEtAl98} (that was used in that work to identify the critical value of $\pi K_0$) is not very pronounced. 

As has been pointed out in Sec.~\ref{Sec:Introduction}, we expect that our results should be relevant to the SIT not only in JJ chains but also in a broader class of 1D systems. In view of this, we briefly discuss also the experimental results for  the SIT in semiconductor nanowires. The theoretical description of such  systems and its mapping to the present model is briefly discussed in  Appendix~\ref{App:supercond-wires}. The SIT was studied  in MoGe nanowires in Refs.~\onlinecite{Bezryadin00,Bollinger08}. While in those works the nanowires were relatively short (with the maximal length 0.5\:$\mu$m), in a later paper\cite{Rogachev12} the transition was analyzed on considerably longer wires (up to 25\:$\mu$m) favorable for the investigation of the infrared physics.  It was found in Ref.~\onlinecite{Rogachev12} that, when the wire cross-section is made smaller, the system undergoes a transition from the superconducting to the insulating phase that is visualized by the behavior of the resistivity $\rho(T)$ with lowering temperature. Also, Ref.~\onlinecite{Rogachev12} demonstrated that application of the magnetic field serves as an alternative way to drive the transition. On this qualitative level, these observations agree with theoretical expectations. A surprising finding of Ref.~\onlinecite{Rogachev12} is that the separatrix curve $\rho(T)$ separating the superconducting and insulating phases is essentially temperature-independent. This is in a clear disagreement with the theoretical expectation of the linear (with a  logarithmic correction) vanishing of the resistivity at the critical line at low temperature, Eq.~(\ref{rho-crit}). This discrepancy might possibly be attributed to the fact that the temperature range in which the resistivity was measured in Ref.~\onlinecite{Rogachev12}, from 2-4\:K down to 0.4\:K was insufficient to probe the infrared asymptotic behavior. An alternative possibility is that some coupling to the environment may have affected the results by suppressing quantum coherence and thus stabilizing the metallic behavior in wires that would otherwise experience an RG flow towards the insulating fixed point.

In a recent preprint \cite{Duty17}, an experimental study of the depinning in JJ chains deeply in the localized phase has been carried out. The authors have found an agreement with theoretical expectations based on the Luttinger-liquid picture in the presence of disorder, which is in correspondence with our model. They have also suggested that previous results on SIT in JJ chains\cite{ChowEtAl98} may have been influenced by an external noise. 

Summarizing this brief discussion of related experiments, we conclude that both classes of 1D systems, JJ arrays and semiconductor nanowires, serve as an outstanding playground for the experimental investigation of SIT in 1D systems. On the other hand, more experimental work is clearly needed to investigate the $T$ and $N$ dependence of resistivity (or resistance) around the SIT and to identify various scaling regimes and the actual position of the transition. 

Before closing the paper, we make comments on two issues that have been left apart in the paper. 

\begin{itemize}

\item[(i)] We have not included random spatial fluctuations of the Luttinger-liquid constant $K_0$ in our effective model. In principle, such fluctuations will also arise as a result of junction-to-junction fluctuations in charging and Josephson energies. We do not expect any essential modifications of our results due to such fluctuations, assuming their relative magnitude is small. On the other hand, this type of disorder may affect essentially the energy transport in a system, since it tends to localize the bosonic modes. A related problem has been considered in the context of quantum wires in Ref.~\onlinecite{Fazio98}. 

\item[(ii)] In the insulating phase, the $\rho(T)$ curves enter, at a certain temperature, the strong-coupling regime (as indicated by dashed lines in our figures).  An interesting question is the fate of the $\rho(T)$ dependences below this temperature. It is expected that the interaction-induced dephasing in a disordered Luttinger liquid\cite{gornyi05a} breaks down in this regime, and the system undergoes, at a non-zero temperature, a many-body localization transition into a phase with infinite resistivity\cite{gornyi05,basko06} (for a recent review see Ref.~\onlinecite{nandkishore15}). This is a true transition only if the coupling to an external bath is zero; its experimental observation thus requires that this coupling is sufficiently weak. The MBL behavior has been experimentally demonstrated in the vicinity of SIT in a disordered 2D system\cite{Ovadia15}. It is natural to expect that it can be observed in the 1D counterpart of such systems as well. 

\end{itemize}

\section{Acknowledgments}

We would like to thank  D.~B. Haviland and A.~D. Zaikin for discussions. A.~D.~M. is grateful to the Weizmann Institute of Science (Weston Visiting Professorship). This work was supported by Russian Science Foundation under Grant No.\ 14-42-00044. We also acknowledge financial support by DFG. 

\begin{appendix}

\section{Derivation of the field theory in the non-local case}
\label{App:Derivation-theory}
 %%%%%%%%%%%%%%%%%%%%%%%%%%%%%%%%%%%%%
\begin{figure}
\centering
\includegraphics[scale=0.27]{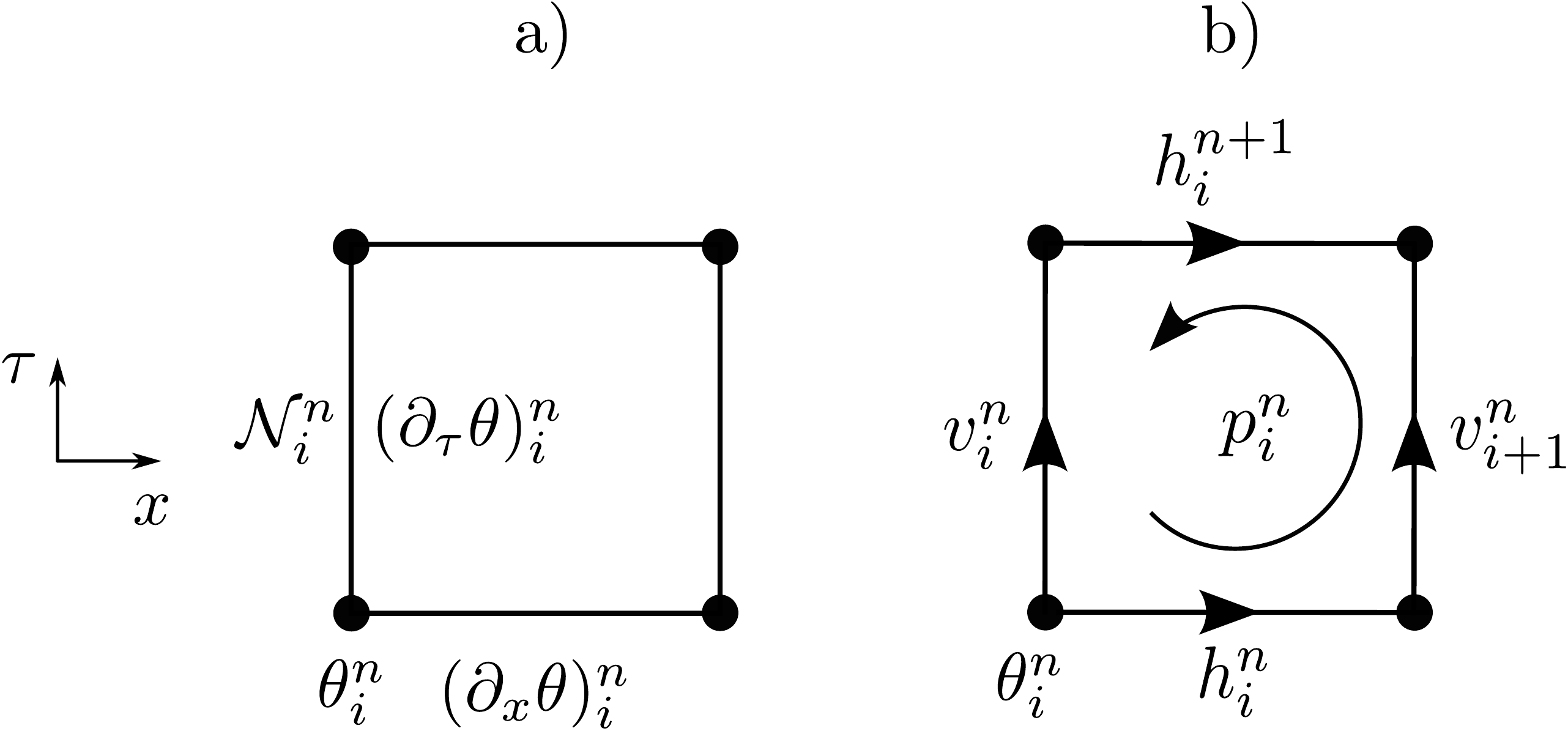}
\caption{Attribution of lattice variables in the derivation of the sine-Gordon theory before (left) and after (right)  the Poisson resummation over charges and of the Villain approximation.}
\label{Fig:variables}
\end{figure}
%%%%%%%%%%%%%%%%%%%%%%%%%%%%%%%%%%%%%%

In this Appendix, we  derive the field-theory description of a JJ chain with arbitrary range of charge-charge interaction. While the derivation follows closely the general procedure outlined in the literature\cite{JoseEtAl77,Minnhagen87,BobbertEtAl90,BenfattoEtAl12,BradleyDoniach84,ChoiEtAl98}, we include it to make the presentation in this paper self-contained. To render the infrared singularities finite, we assume the system to have ring geometry with $N_x$ islands and junctions.  We focus here on the case of a clean JJ array and derive Eqs. (\ref{Eq:S0}) and (\ref{Eq:Sps}) of the main text. The generalization to the case of a disordered chain and the derivation of Eq. (\ref{Eq:SpsQ}) is then straightforward.  

We aim at deriving the path-integral formulation of the partition function of the system. To this end, we discretize the imaginary time into $N_{\tau}$ steps. The step size $\Delta \tau$ is chosen to be of the order of the characteristic time for the local dynamics in the JJ chain,
\begin{equation}
\Delta\tau=\sqrt{\frac{E_1+E_0}{\Ej E_1 E_0}} = \frac{1}{\Omega_0}.
\label{Eq:A:DeltaTau}
\end{equation} 
The time step (\ref{Eq:A:DeltaTau}) interpolates between $1/\sqrt{\Ej E_0}$ in the local-interaction limit and $1/\sqrt{\Ej E_1}$ in the limit $\Lambda\rightarrow\infty$. 
 
At each vertex of the space-time lattice obtained after the discretization,  we  introduce  a resolution of identity,
\begin{equation}
\mathbb{1}=\sum_{\cal N}\int_{-\pi}^{\pi}\dfrac{\diff \theta}{2\pi}\left|{\cal N}\right>\left<\theta\right|\e^{- i  \theta {\cal N}},
\end{equation}
where $\theta$ is the phase  of a superconducting island and $\mathcal{N}$ is its  charge. 
In what follows, we attribute the phase $\theta_i^n$ to the site $(x,\tau)=(i,n)$   and  the island charges ${\cal N}_i^n$ to the vertical links of the space-time lattice. The components of the (discrete) gradients of the $\theta$-field are denoted by $\partial_x\theta$ and $\partial_{\tau}\theta$  and are attributed to the horizontal and vertical links, respectively. We summarize our notations in Fig.~\ref{Fig:variables} (left panel).

The imaginary-time action describing our system assumes now the form:
\begin{widetext}
\begin{equation}
S=- i  \sum_{\mathrm{vert. links}}{\cal N}_i^n\left(\partial_{\tau}\theta\right)_i^n+\frac{E_{1}\Delta \tau}{2}\sum_{\mathrm{vert. links}}S_{ij}^{-1} {\cal N}_i^n {\cal N}_j^n+\Ej \Delta\tau \sum_{\mathrm{hor. links}}\left(1-\cos\left[\left(\partial_x\theta\right)_i^n\right]\right).
\end{equation}
\end{widetext}
We proceed by performing Poisson resummation over the charges ${\cal N}_i^n$ in favor of a new integer-valued field $v_i^n$ (also defined on vertical links). 
Furthermore, we adopt the Villain approximation for the Josephson couplings 
\begin{equation}
\exp[-\Ej\Delta\tau (1-\cos\gamma)]\approx\sum_{h}\exp\left[-\frac{\Ej\Delta\tau}{2}(\gamma+2\pi h)^2\right]
\end{equation} 
and get 
\begin{widetext} 
\begin{equation}
S=\frac{K_1^2}{2K}\sum_{\rm vert. links }S_{ij}
\big[\left(\partial_\tau \theta\right)^n_i -2\pi v^{n}_i\big]
\left[\left(\partial_\tau \theta\right)^n_j -2\pi v^{n}_j\right] 
+\frac{K}{2} \sum_{\rm hor. links }
\left[\left(\partial_x \theta\right)^n_i -2\pi h^{n}_i\right]^2, 
\label{Eq:A:Sthvh}
\end{equation}
\end{widetext}
where $K=\sqrt{\Ej(E_1+E_0)/E_1E_0}$
and $h_i^n$ is an integer-valued field attributed to the horizontal links. 
 The partition function of the model now reads
 \begin{equation}
 Z=\int_0^{2\pi} D\theta \sum_{\{v\}, \{h\}}e^{-S}.
 \label{Eq:A:Zthvh}
 \end{equation}
 Note  that the integration over each $\theta^n_i$ in Eq. (\ref{Eq:A:Zthvh}) is limited to the interval $(0, \, 2\pi)$.
 However, the summation over the longitudinal (with zero curl) part of the vector field   $(h^n_i, v^n_i)$ promotes the integration over the superconducting 
 phase to the full real line, resulting in an ordinary Gaussian integral.  
 Thus, eliminating $\theta_i^n$ and the longitudinal component of  $(h^n_i, v^n_i)$ from the partition function we get the action for the ``vorticity''  of the vector field~$(h^n_i, v^n_i)$. The latter is characterized by the circulation of  $(h^n_i, v^n_i)$ around each elementary plaquette of our lattice (see right panel of Fig. \ref{Fig:variables}),
 \begin{equation}
 p_i^n=h_i^n+v_{i+1}^n-h_i^{n+1}-v_i^n,
 \end{equation}
 together with circulations over the two global loops in the system,
 \begin{equation}
 H_0=\sum_i h_i^0 \qquad \mathrm{and} \qquad V_0=\sum_n v_0^n.
 \label{Eq:A:H0V0}
 \end{equation}

In terms of vorticities introduced above  and in Fourier space (with dimensionless frequency $\omega$) the action of our model acquires the form
 \begin{multline}
S=\frac{2\pi^2}{N_xN_\tau E_0\Delta\tau}\bar{V}_0^2+\frac{2\pi^2 \Ej \Delta \tau}{N_xN_\tau }\bar{H}_0^2\\+\frac{2\pi^2K}{N_xN_\tau}
\sum_{(q,\w)\neq 0}U^{-1}(\omega, q)|p(\omega, q)|^2,
\label{Eq:A:SVFull}
\end{multline}
 where 
 \begin{eqnarray}
 U(\omega, q)= \Delta(\omega)+\frac{\Delta(q)}{(1-\ug)\Delta(q)+\ug}, \label{Eq:A:U1}\\
 \Delta(\xi)=2(1-\cos\xi)\,, \qquad \ug=\frac{1}{1+\Lambda^2}, \label{Eq:A:U2}
 \end{eqnarray}
 and
 \begin{eqnarray}
\bar{V}_0&=&N_x V_0+\sum_{i=1}^{N_x-1}i\sum_n p_i^n, \quad\\
\bar{H}_0&=&N_{\tau} H_0-\sum_{n=1}^{N_{\tau}-1}n\sum_i p_i^n.\quad
\end{eqnarray}
 
Applying the Poisson resummation procedure to the summation over $V_0$ in the partition function, one can see that this summation is equivalent to the summation over all possible sectors of the theory with different total charge of the chain and can be safely dropped. 
Straightforward algebra allows then to rewrite the action (\ref{Eq:A:SVFull}) as 
\begin{equation}
S=\frac{2\pi^2 \Ej\Delta\tau}{N_x}\sum_{n=0}^{N_\tau-1} H_n^2+
 \frac{2\pi^2K}{N_x N_\tau}\sum_{q\neq 0,\omega }
U^{-1}(\omega, q)|p(\omega, q)|^2,
\label{Eq:A:SFinal}
\end{equation}
where in the last sum all the terms with $q=0$ are excluded and  [cf. definition (\ref{Eq:A:H0V0})]
\begin{equation}
 H_n=\sum_i h_i^n. 
\end{equation}

We are now in a position to derive the sine-Gordon type description of our system. 
First, we introduce the Hubbard-Stratonovich field $\tilde{\phi}$ and decouple the vortex interaction term in Eq. (\ref{Eq:A:SFinal}) according to
\begin{widetext}
\begin{equation}
\exp\left\{\!\!- \frac{2\pi^2K} {N_x N_\tau}\!\!\sum_{q\neq 0,\w}\!\!U^{-1}(\omega, q)|p(\omega, q)|^2\!\!\right\}\!\propto\!\! \int\!\! {\cal D}\tilde{\phi}\exp\left\{\!\!-\frac{1}{2\pi^2 K N_xN_\tau}\!\!\sum_{q\neq 0,\w}U(\omega, q)|\tilde{\phi}(\omega, q)|^2\!+\!\frac{2 i }{N_xN_\tau}\!\!\sum_{q\neq 0,\w} \!\!\tilde{\phi}(\omega, q)p^*(\omega, q)\!\!\right\}.
\end{equation}
\end{widetext}
 Note that the field $\tilde{\phi}(x, \tau)$ introduced here by definition has no $q=0$ Fourier components.  
 However, the variables  $H_n$ in 
 Eq. (\ref{Eq:A:SFinal})  are related to the local vorticities $p_n^i$ via $H^{n}-H^{n+1}=\sum_i p_i^n$. In order to carry out the summation over $H_n$ in the partition function, one thus needs to introduce an additional field $\phi_0(\tau)$ resolving the corresponding Kronecker $\delta$-function constraint,
 \begin{multline}
 \delta\left(H_{n}-H_{n+1}-\sum_i p_i^n\right) \\=\int_0^\pi \frac{d\phi_0(n)}{\pi} \exp\left[-2 i \phi_0(n)\left(H^{n}-H^{n+1}-\sum_i p_i^n\right)\right].
 \label{Eq:Kronecker-delta}
 \end{multline} 
 The summations over $H_n$ lead now to the action
 \begin{equation}
 S=\frac{1}{2\pi^2 K N_xN_\tau}\!\sum_{q, \w}\!U(\omega, q)|\phi(\omega, q)|^2+2 i \!\sum_{x, \tau}\!\phi(x, \tau)p(x, \tau),
 \label{Eq:A:Sphip}
 \end{equation}
 where the field $\phi$ is compact in $\tau$-direction (with compactification radius $\pi$) and possesses the mode expansion
\begin{equation}
\phi(x, \tau)=\phi_0(\tau)+\frac{\pi m \tau}{N_\tau}+\frac{1}{N_x}\sum_{q\neq 0}\tilde{\phi}_q(\tau).
\end{equation}

All the transformations performed so far  on our lattice model were essentially exact (up to Villain approximation).  However, our treatment misses the physics at time scales shorter when $\Delta\tau$ that is the characteristic time for a quantum phase slip.  It is expected on physical grounds that quantum phase slips described by the vortex numbers $p(x, \tau)$ bear some action cost $S_{\rm short}$ (per phase slip) coming from those omitted time scales. Therefore, the action 
(\ref{Eq:A:Sphip}) should be modified by adding a correction term 
\begin{equation}
\delta S=S_{\rm short}\sum_{x, \tau}p^2(x, \tau).
\end{equation}
When the superconducting correlations are (locally) well developed, $S_{\rm short}\gg 1$,  we can perform the summation over $p(x, \tau)$ to the lowest order 
in $y=\exp[-S_{\rm short}]$, the amplitude of a quantum phase slip. The resulting action takes the form of a sine-Gordon theory:
 \begin{equation}
 S=\frac{1}{2\pi^2 K N_xN_\tau}\sum_{q, \w}U(\omega, q)|\phi(\omega, q)|^2+2y\sum_{x, \tau}\cos 2\phi(x, \tau).
 \label{Eq:A:SFinalDiscreet}
 \end{equation}
Here $U(\omega, q)$ is given by Eqs. (\ref{Eq:A:U1}) and (\ref{Eq:A:U2}).

Equation (\ref{Eq:A:SFinalDiscreet}) constitutes the main result of this Appendix. In the continuum limit, $q,\omega \ll 1$,  it reduces to
\begin{multline}
 S=\frac{1}{2\pi^2 K}\int_{-1}^1\frac{dq}{2\pi}\int_{-\Omega_0}^{\Omega_0}\frac{d\omega}{2\pi}U(\omega, q)|\phi(\omega, q)|^2\\+2y\Omega_0 \int dx d\tau 
 \cos 2\phi(x, \tau) \,,
 \label{Eq:A:SFinalCont}
 \end{multline} 
where we have restored the physical dimension of frequency and 
\begin{eqnarray}
& \displaystyle U(\omega, q)= \frac{\omega^2}{\Omega_0}+\frac{q^2\Omega_0}{(1-\ug)q^2+\ug}, \quad \ug=\frac{1}{1+\Lambda^2},\qquad \label{Eq:A:U1Final}\\[0.2cm]
& \displaystyle K=\sqrt{\frac{\Ej(E_1+E_0)}{E_1E_0}}, \quad\Omega_0=\sqrt{\frac{\Ej E_1 E_0}{E_1+E_0}}.  
 \label{Eq:A:U2Final} 
\end{eqnarray}
Equations (\ref{Eq:A:SFinalCont}), (\ref{Eq:A:U1Final}), and (\ref{Eq:A:U2Final}) are equivalent to Eqs.~(\ref{Eq:S0}) and (\ref{Eq:Sps}) of the main text.   

Before closing this Appendix, let us briefly mention another justification of the transformation of  the action (\ref{Eq:A:Sphip}) to the form (\ref{Eq:A:SFinalCont}). 
To this end, we note that Eq. (\ref{Eq:A:SFinalCont}) with $y\sim 1$ can be understood as the first term in the expansion of the effective action for the field $\phi$ 
in harmonics $\cos n\phi$. On the other hand, the RG equations for the action  (\ref{Eq:A:Sphip}) discussed in the main text show that the QPS amplitude $y$ rapidly renormalizes down on the first few steps of the RG procedure provided that  the system is locally superconducting and that at short scales $K\gg 1$. It is also easy to show that the amplitudes of higher harmonics vanish even faster. 
Thus, Eq. (\ref{Eq:A:SFinalCont}) constitutes an adequate description of the system on length scales larger than the lattice spacing and time scales larger then $\Delta\tau$. A further discussion of this point can be found in Appendix \ref{A:MLG} where the action (\ref{Eq:A:Sphip}) is analyzed in full detail for the case of infinite-range Coulomb interaction, $\Lambda=\infty$.

\section{Infinite-range interaction}
\label{A:MLG}

In this Appendix, we  study the theory in the limit of infinite-range Coulomb interaction ($\Lambda \to \infty$). For this special case we provide another connection between our lattice model and the sine-Gordon theory (which supports the results of Apendix \ref{App:Derivation-theory}), derive an estimate for the fugacity, Eq.~(\ref{Eq:y}), and  compare our results  to previous works.

Our starting point is Eq.~\eqref{Eq:A:SFinal}. In the limit $\Lambda \to \infty$ ($u_{\rm{g}}\to 0$), the interaction between vorticities $U^{-1}(\w,q)$ is momentum independent and gapped. We thus approximately find
\begin{eqnarray}
\frac{1}{V}\!\!\! \sum_{q\neq 0,\w\!\!\!} U^{-1}(\w,q) |p(\w,q)|^2 & \simeq & \frac{1}{V}\!\!\sum_{q\neq 0,\w}\!\!|p(\w,q)|^2 \nonumber
\\
&= & \sum_{n,i}\left(p_i^n\right)^2\!-\!\frac{1}{N_x} \sum_{n}\!\left[\sum_i p_i^n\right]^2.  \nonumber \\
\end{eqnarray} 
In the next step we perform the summations over vorticities $\{p\}$. As discussed in Appendix \ref{App:Derivation-theory}, we incorporate the constraint $H^n-H^{n+1}=\sum_i p_i^n$ via the introduction of an auxiliary field $\phi^n \in (0,\pi )$ [cf. Eq.~\eqref{Eq:Kronecker-delta}]. We then arrive at the action
\begin{eqnarray}
S&=&\frac{(2\pi)^2K_1}{2N_x}\sum_n (H^n)^2+2 i \sum_n \phi^n \left(H^{n+1}-H^n\right)  \nonumber \\
&-& \Delta \tau\sum_nU(\phi^n),
\label{Eq:App:action_with_potential}
\end{eqnarray}
where
\begin{equation}
\begin{split}
\e^{-\frac{U(\phi)}{\sqrt{E_{\rm{J}}E_1}}}&=\sum_{z}\e^{2  i  \phi z} \exp\left[-f(z)\right],
\\
f(z)&=-\frac{(2\pi)^2K_1}{2N_x}z^2-\ln \int_{0}^{\pi} \frac{d\omega}{\pi}e^{2 i \omega z}(g(\omega))^{N_x},
\\
g(\omega)&=\sum_{p}\exp\left[-\frac{K_1(2\pi)^2p^2}{2}-2 i \omega p\right].
\end{split}
\end{equation}
The action \eqref{Eq:App:action_with_potential} describes a particle on a ring (coordinate $\phi$ and momentum $H$) moving in a potential $U(\phi)$. Since $f(z)$ is a periodic function with period $N_x$, we can write
\begin{equation}
\begin{split}
\e^{-\frac{U(\phi)}{\sqrt{E_{\rm{J}}E_1}}}&=\sum_{z_0=0}^{N_x-1}\e^{2 i  \phi z_0}\exp[-f(z_0)]\sum_{z_1}\e^{2 i  N_x z_1 \phi}
\\
&=\frac{\pi}{N_x}\sum_{z_0=1}^{N_x-1}\e^{2 i  \phi z_0}\exp[-f(z_0)]\sum_{k}\delta\left(\phi-\frac{\pi k}{N_x}\right).
\end{split}
\end{equation}
We now observe that the potential $U(\phi)$ is not a smooth function. However, we argue that for $N_x\gg 1$, the exponential $\exp\left[U(\phi)/\sqrt{\Ej E_1}\right]$ converges in the sense of distributions to the discrete Fourier transform
\begin{equation}
\frac{1}{\pi} \exp\left[-\frac{U(\phi)}{\sqrt{\Ej E_1}}\right]\to\sum_{z_0=0}^{N_x-1} \e^{2 i  \phi z_0} \exp[-f(z_0)].
\end{equation}
One can show now  that for  $N_x\gg1 $ and $K_1 \gg 1$ the potential  $U(\phi)$ is approximately given by
\begin{equation}
U(\phi) =2 y(N_x) [1-\cos2\phi]+\mathrm{const} \,,
\label{U-phi}
\end{equation}
where
\begin{equation}
y(N_x) =\sqrt{\Ej E_1}N_x \exp\left[-2\pi^2 K_1 \left(1-\frac{1}{N_x}\right)\right].
\label{A:Eq:yInfiniteRange}
\end{equation}

We have thus reduced the model of a JJ chain with infinite-range interaction of a length $N_x$ to a quantum mechanics with the cosine potential (\ref{U-phi}), i.e., to a zero-dimensional version of the sine-Gordon theory. Equation  (\ref{A:Eq:yInfiniteRange}) thus yields the  QPS amplitude at scale  $N_x$ for the chain with infinite-range interaction. 
These results can be compared to the findings of Refs.  \onlinecite{MatveevEtAl02,RastelliEtAl13}  where the suppression of the persistent current by QPS in a JJ chain with ring geometry was studied. It was found there that the effective QPS amplitude at length $N_x$ is given by
\begin{equation}
y(N_x) \propto \Ej^{3/4} E_1^{1/4}N_x\e^{-8 K_1 (1-\gamma/N_x)} \,,
\label{A:Eq:yMLG}
\end{equation}
with $\gamma =1/2+\pi^2/8$. Comparing Eqs. (\ref{A:Eq:yInfiniteRange})  and (\ref{A:Eq:yMLG}), we see that our approach yields the same form of the QPS amplitude in its dependence on $K_1$ and $N_x$ as was obtained in Ref.~\onlinecite{MatveevEtAl02,RastelliEtAl13}. On the other hand, the numerical coefficient in the exponent of Eq. (\ref{A:Eq:yInfiniteRange}) is different from that in Eq.~(\ref{A:Eq:yMLG}).  This difference in the numerical coefficient can be traced back to the fact that our model is only approximate at the scale of the order of the bare ultraviolet cutoff of the problem ($N_{x}\sim 1$). 

%%%%%%%%%%%%%%%%%%%%%%%%%%%%%%%%%%%%%%%%%%%%%%%%%%%%%%%%%%%%%%%%%%%%%%%%%%%%%%%%%%%%%%%%%%%%%%%%%%%%%%%%%%%%%%%%%%%%%%%%%%%%%%%%%%%%%%%%%%%%%%%%%%%%%%%%%

\section{Derivation of RG equations for short-range interaction}
\label{App:RG-short_range}

In this Appendix, we present the main steps of the derivation of the RG equations in the case of short-range Coulomb interaction, 
Eqs.~(\ref{Eq:RG_K0_local})--(\ref{Eq:rD}) of Sec.~\ref{Sub:RG-short-range}. We calculate the correlation function \eqref{Eq:correlation_function_R} perturbatively in $y$ ($2^{\mathrm{nd}}$ order) and $D_{\xi}$ ($1^{\mathrm{st}}$ order). To this order, phase slips and disorder do not mix. The correction due to disorder ($\propto  D_{\xi}$) is therefore the same as in Ref.~\onlinecite{GiamarchiSchulz88}. In zeroth order we obtain
\begin{equation}
R^{(0)}(\mathbf{r}) =  \e^{-2\pi K_0 F_1(\mathbf{r})},
\ee
where  
\be 
F_1(x,\tau) = \frac{1}{2}\ln\left(\frac{x^2+(u_0|\tau|+a)^2}{a^2}\right)\,.
\ee
Taking into account the contribution to the phase slips yields 
\bea
R(\mathbf{r}) &=&  \e^{-2\pi K_0 \tilde F_1(\mathbf{r})},  \\  
\tilde{F}_1(x,\tau) &=& F_1(x,\tau) + \frac{d}{K_0}\cos\left(2\theta_{\mathbf{r}}\right)\,,
\eea
where $\theta_{\mathbf{r}}$ is the angle between the vector $(x,u_0\tau)$ and the $x$-axis. The constant $d$ parametrizes the anisotropy between space and time \cite{GiamarchiSchulz88}. Initially $d=0$ but it gets generated during the RG. For the purpose of this derivation we explicitly reintroduce the lattice spacing $a$. In the following we calculate the second-order correction $\propto y^2$. To this end, we exploit the following equality for the averaging over the Gaussian action of the clean system:
\begin{widetext}
\begin{equation}
\begin{split}
\lim_{n\to 0}\sum_{a=1}^n&\left< \e^{2 i  [\phi^j(\mathbf{r}_1)-\phi^j(\mathbf{r}_2)]}  \cos\left[2\phi^a(\mathbf{r}_1)\right]\cos\left[2\phi^a(\mathbf{r}_2)\right] \right>_0 
\\
&\qquad \qquad \qquad 
= \frac{1}{4}\e^{-2\pi K_0[F_1(\mathbf{r}_1-\mathbf{r}_2)+F_1(\mathbf{r}_3-\mathbf{r}_4)]}\sum_{\sigma=\pm} \left[\e^{2\pi K_0\sigma[F_1(\mathbf{r}_1-\mathbf{r}_3)+F_1(\mathbf{r}_2-\mathbf{r}_4)-F_1(\mathbf{r}_1-\mathbf{r}_4)-F_1(\mathbf{r}_2-\mathbf{r}_3)]}-1 \right].
\end{split}
\end{equation}
The average over the random stray charges is evaluated by assuming a Gaussian distribution,
\begin{equation}
P[Q]=\exp\left\{-\frac{\pi^2}{D_{Q}a}\int \diff x\, Q^2(x)\right\}.
\end{equation}
We find
\begin{equation}
\left<\cos[{\cal{Q}}(x_3)]\cos[{\cal{Q}}(x_4)]+\sin[{\cal{Q}}(x_3)]\sin[{\cal{Q}}(x_4)] \right>_{Q}=\exp\left\{-D_{Q}\frac{|x_3-x_4|}{a}\right\}.
\end{equation}
The second-order correction to $R$ assumes thus the form [here $\mathbf{r}_i=(x_i,u_0 \tau_i)$]:
\begin{equation}
\frac{y^2}{16\pi^3a^4}\e^{-2\pi K_0 F_1(\mathbf{r}_1-\mathbf{r}_2)}\!\!\int \!\!\diff^2 r_3 \diff^2 r_4 \e^{-2\pi K_0 F_1(\mathbf{r}_3-\mathbf{r}_4)}\e^{-D_{Q} \frac{|x_3-x_4|}{a}}\!\sum_{\sigma=\pm} \left[\e^{2\pi K_0\sigma[F_1(\mathbf{r}_1-\mathbf{r}_3)+F_1(\mathbf{r}_2-\mathbf{r}_4)-F_1(\mathbf{r}_1-\mathbf{r}_4)-F_1(\mathbf{r}_2-\mathbf{r}_3)]}-1 \right].
\end{equation}
The first exponential factor in the above integrand is a power-law function of $r=|\mathbf{r}_3-\mathbf{r}_4|$. This allows us to expand the square bracket in $r$ and perform the integration over the polar angle of $\mathbf{r}$. While carrying out the integration over the center mass coordinate $\mathbf{R}=(\mathbf{r}_3+\mathbf{r}_4)/2$, we use the following identities
\begin{align}
\int \diff^2 R \,[F_1(\mathbf{R}-\mathbf{r}_1)-F_1(\mathbf{R}-\mathbf{r}_2)]\left(\partial_X^2+\partial_Y^2\right)[F_1(\mathbf{R}-\mathbf{r}_1)-F_1(\mathbf{R}-\mathbf{r}_2)]&=-4\pi F_1(\mathbf{r}_1-\mathbf{r}_2),
\\
\int \diff^2 R \,[F_1(\mathbf{R}-\mathbf{r}_1)-F_1(\mathbf{R}-\mathbf{r}_2)]\left(\partial_X^2-\partial_Y^2\right)[F_1(\mathbf{R}-\mathbf{r}_1)-F_1(\mathbf{R}-\mathbf{r}_2)]&=-2\pi\cos 2\theta_{\mathbf{r}_1-\mathbf{r}_2}.
\end{align}
As a result, we obtain for the second-order correction
\begin{equation}
2\pi\frac{y^2}{2}K_0^2\!\int_a^{\infty}\!\frac{\diff r}{a}\left(\frac{r}{a} \right)^{3-2\pi K_0}\!\!\Biggl\{\! F_1(\mathbf{r}_1-\mathbf{r}_2)\! \left[ \BesselI_0\left(\! D_{Q} \frac{r}{a}\right)-\StruveL_0\left(\! D_{Q} \frac{r}{a}\right)\right]
+\frac{1}{2}\cos2\theta_{\mathbf{r}_1-\mathbf{r}_2}\!\left[\BesselI_2\left(\!D_Q \frac{r}{a}\right)-\StruveL_2\left(\!D_Q \frac{r}{a}\right)-\frac{2}{3\pi}D_Q \frac{r}{a}\right]\!\Biggr\}\!,
\label{RG-corrections}
\end{equation} 
where $\BesselI_n$ are modified Bessel functions of the first kind and $\StruveL_n$ are modified Struve functions. 
Equation (\ref{RG-corrections}) represent starting terms of the expansion of $\exp\left[-2\pi K_0^{\rm{eff}} \tilde{F}_1(\bf r_1- \bf r_2) \right]$ with
\begin{align}
K_0^{\rm{eff}}&=K_0-\frac{y^2}{2}K_0^2\int_a^{\infty}\frac{\diff r}{a}\left(\frac{r}{a} \right)^{3-2\pi K_0}\left[ \BesselI_0\left(D_{Q}  \frac{r}{a}\right)-\StruveL_0\left( D_{Q} \frac{r}{a}\right)\right],
\\
d_{\rm{eff}}&=d-\frac{y^2}{4}K_0^2\int_a^{\infty}\frac{\diff r}{a}\left(\frac{r}{a} \right)^{3-2\pi K_0}\left[\BesselI_2\left(D_Q \frac{r}{a}\right)-\StruveL_2\left(D_Q \frac{r}{a}\right)-\frac{2}{3\pi}D_Q \frac{r}{a}\right],
\end{align}
and 
\be
\tilde{F}_1({\bf r}) = F_1({\bf r}) + \frac{d_{\rm{eff}}}{K_0^{\rm{eff}}}\cos\left(2\theta_{\mathbf{r}}\right)\,.
\ee
The effective constants $K_0^{\rm{eff}}$ and $d_{\rm{eff}}$ determine the low-energy behavior of the correlator $R$. Hence, varying the cutoff $a \to a+\diff a$ should not change them. Consequently, we find
\begin{align}
K_0(a+\diff a)&=K_0(a)-\frac{y^2}{2}K_0^2\left[ \BesselI_0\left( D_{Q} \right)-\StruveL_0\left( D_{Q}  \right)\right]\frac{\diff a}{a},
\\
d(a+\diff a)&=d(a)-\frac{y^2}{4}K_0^2\left[\BesselI_2\left(D_Q \frac{r}{a}\right)-\StruveL_2\left(D_Q \frac{r}{a}\right)-\frac{2}{3\pi}D_Q \frac{r}{a}\right]\frac{\diff a}{a},
\\
y^2(a+\diff a)&=y^2(a)\left(\frac{a+da}{a}\right)^{4-2\pi K_0},
\\
D_{Q}(a+da)&=D_{Q}(a)\frac{a+da}{a}.
\end{align}
\end{widetext}
Using the parametrization $a(l)=\e^l$ results in the RG equations for $K_0$, $y$ and $D_Q$ stated in the main text in Sec.~\ref{Sub:RG-short-range}, see Eqs.~(\ref{Eq:RG_K0_local})--(\ref{Eq:rD}). The relation between the renormalization of $d$ and $u_0$ is given by \cite{GiamarchiSchulz88}
\begin{equation}
\frac{\diff u_0}{\diff l}=-2\frac{u_0}{K_0}\frac{\diff d}{\diff l}.
\end{equation}
The contribution of $S_{\xi}$ to renormalization (i.e., terms linear in $D_\xi$) is the same as in Ref.~\onlinecite{GiamarchiSchulz88}.

\section{Memory function}
\label{App:memory_function}

In this Appendix, we present the calculation of the memory function in the limit of local Coulomb interaction, Sec.~\ref{Sub:Transport-short-range}. We use the action \eqref{Eq:Full-action} in the local limit $\Lambda \to 0$ and go over to the Hamiltonian description:
\begin{align}
\Ham&=\Ham_0+\Ham_{\rm{ps}}+\Ham_{\xi},
\\
\Ham_0&=\frac{1}{2}\int\diff x\,\left[u_0K_0\left(\partial_x\theta\right)^2+\frac{u_0}{\pi^2K_0}\left(\partial_x \phi\right)^2\right],
\\
\Ham_{\rm{ps},Q}&=\frac{y u_0}{\sqrt{2\pi^3}a^2}\int \diff x\,\cos\left[2\phi(x)- {\cal{Q}}(x)\right],
\\
\Ham_{\xi}&=\int\diff x \left[ \frac{\xi(x)}{a^{3/2}} \e^{2 i  \phi(x)}+ \mathrm{h. c.}\right].
\end{align}
Here, we have again explicitly introduced the lattice spacing $a$.
The commutator of $\Ham$ with the current operator splits into two parts: $F=F_{\rm{ps}}+F_{\xi}$, where
\begin{align}
\begin{split}
F_{\rm{ps}}&=-2\sqrt{\frac{2}{\pi}} i  e u_0^2 K_0 \frac{y}{a^2} \Bigl[\sin 2\phi(x) \cos{\cal Q}(x)
\\
&\qquad \qquad \qquad \qquad \quad + \cos 2\phi(x) \sin {\cal Q}(x) \Bigr],
\end{split}
\\
F_{\xi}&=-4\pi e \frac{u_0}{a^{3/2}} K_0 \left[\xi(x) \e^{2 i  \phi(x)}-\xi^{\ast}(x) \e^{-2 i  \phi(x)} \right].
\end{align}
In order to proceed, we have to compute the correlation function 
\begin{equation}
C(\w)=\int \diff x \int_0^{\infty} \diff t\, \e^{ i  \w t} \left< [F(x,t),F(0,0)] \right> \,,
\end{equation}
where the  angular brackets denote averaging over disorder as well as with respect to $\Ham$. To the lowest order in $y$ and $D_{\xi}$, the averaging can be  performed with respect to $\Ham_0$ instead of the full Hamiltonian. To this order, the correlation function $C(\w)$ splits into two independent parts: $C=C_{\rm{ps}}+C_{\xi}$. In the following, we sketch the calculation of both of them.

\subsection{Random fugacity part}

The conductivity of a disordered 1D system has  been calculated in Ref.~\onlinecite{LiOrignac02} using the memory function formalism. For completeness, we demonstrate here the key steps.
We calculate the (time-ordered) correlation function 
\begin{equation}
C_{\xi}(x,\tau)=\left<T_{\tau} F_{\xi}(x,\tau) F_{\xi}(0,0) \right>
\end{equation}
in imaginary time $\tau$ and analytically continue it  to real time afterwards. 
The quantum average over exponentials yields 
\begin{equation}
\left<\e^{2 i  \phi(x,\tau)}\e^{-2 i  \phi(0,0)}\right>_{0}=\e^{-2\pi K_0 F_1(x,\tau)},
\end{equation} 
where at finite temperature $1/\beta$
\begin{equation}
F_1(x,\tau)=\frac{1}{2}\ln\left[ \frac{\beta^2 u_0^2}{\pi^2 a^2}\sinh\left(\frac{\pi}{u_0\beta}x_+ \right)\sinh\left(\frac{\pi}{u_0\beta}x_- \right) \right]
\end{equation}
and $x_{\pm}=x\pm  i  u_0 \tau$. After the disorder averaging, we arrive at
\begin{equation}
C_{\xi}(x,\tau)=-8  e^2\frac{u_0^4}{a^3} K_0^2 D_{\xi} \delta(x) \e^{-2\pi K_0 F_1(x,\tau)}.
\end{equation}
The corresponding retarded function can be found via analytical continuation (cf. Ref.~\onlinecite{GiamarchiBook}),
\begin{equation}
C_{\xi}(t>0)=A \left(\frac{\pi a}{u_0 \beta}\right)^{2\pi K_0}\!\left[\sinh\left(\frac{\pi t}{\beta}\right)\right]^{-2\pi K_0}\!,
\end{equation}
where $A=16e^2(u_0^4/a^3)K_0^2D_{\xi}\sin(\pi^2 K_0)$. The Fourier transform to real frequency thus reads
\begin{align}
C_{\xi}(\w)&=\int_{0}^{\infty} \diff t \, \e^{ i  \w t} C_{\xi}(t) \nonumber
\\
&= \frac{A\,a}{u_0} \left(\frac{2\pi a }{\beta u_0} \right)^{2\pi K_0-1}
\mathrm{B}\left(1-2\pi K_0,\pi K_0-\frac{ i  \w \beta}{2\pi}\right),
\label{Cxiw}
\end{align}
where $\mathrm{B}(x,y)$ denotes the Euler Beta function.
The integral in Eq.~(\ref{Cxiw}) converges only for $2\pi K_0<1$ but can be analytically continued to arbitrary $K_0$. The memory function assumes in the limit $\w \to 0$ the form
\begin{equation}
M_{\xi}(T)=\frac{2\pi  i  \,\Gamma^2(\pi K_0) u_0 K_0}{\Gamma(2\pi K_0)a}D_{\xi}\left(\frac{2\pi a T}{u_0}\right)^{2\pi K_0-2} \,,
\end{equation}
which leads to Eq.~(\ref{sigma-xi}) of the main text.

\subsection{Phase-slip part}

For the phase-slip contribution we compute
\begin{align}
C_{\rm{ps}}(x,\tau)&=\left<T_{\tau} F_{\rm{ps}}(x,\tau)F_{\rm{ps}}(0,0) \right>
\\
&=-\frac{4}{\pi} e^2 (u_0/a)^4 K_0^2 y^2 \e^{-2\pi K_0F_1(x,\tau)}\e^{-D_{Q}|x|/a}.
\end{align}
After analytic continuation, we arrive at
\begin{widetext}
\begin{equation}
C_{\rm{ps}}(\w)=\gamma \left(\frac{\pi a}{u_0 \beta}\right)^{2\pi K_0}\int_{0}^{\infty}\diff t \int_{-u_0 t}^{u_0 t} \diff x \frac{e^{ i  \w t}\e^{-D_{Q} |x|/a}}{\left[\sinh\left(\frac{\pi}{u_0\beta}(u_0 t-x)\right)\sinh\left(\frac{\pi}{u_0\beta}(u_0 t+x)\right)\right]^{\pi K_0}},
\end{equation}
where $\gamma=(8/\pi) e^2 (u_0/a)^4 K_0^2 y^2 \sin(\pi^2 K_0)$. It is convenient to change variables to the dimensionless light-cone variables $z=\pi/(u_0\beta) \,(u_0 t+x)$ and $\bar{z}=\pi/(u_0\beta) \,(u_0 t-x)$:
\begin{equation}
C_{\rm{ps}}(\w)=\gamma \frac{u_0 \beta^2}{2\pi^2}\left(\frac{\pi a }{u_0 \beta}\right)^{2\pi K_0}\int_{0}^{\infty}\diff z \int_{0}^{\infty} \diff \bar{z}\, \frac{e^{ i \frac{\beta \w}{2\pi} (z+\bar{z})}\e^{-\frac{D_{Q}u_0\beta}{2\pi a}|z-\bar{z}|}}{\left[\sinh\left(z \right)\sinh\left(\bar{z}\right)\right]^{\pi K_0}}.
\end{equation}
The integrals are convergent for $0<K_0<1/\pi$ but can be analytically continued. We analyze both integrals in the limit of weak and strong disorder. 

For $D_{Q} u_0 \beta/a  \ll 1$, we find in zeroth order
\begin{equation}
C^{(0)}_{\rm{ps}}(\w)=\frac{\gamma}{4} \frac{u_0 \beta^2}{2\pi^2}\left(\frac{2\pi a}{u_0 \beta}\right)^{2\pi K_0}\mathrm{B}^2\!\left(1-\pi K_0,\frac{\pi K_0}{2}-\frac{ i  \beta \w}{4\pi}\right).
\end{equation}
The correction linear in $D_{Q}$ reads in the limit $\w \to 0$
\begin{equation}
C^{(1)}_{\rm{ps}}(\w)\stackrel{\w \to 0}{\to} -\frac{D_{Q} u_0 \beta}{2\pi a}\gamma \frac{u_0 \beta^2}{2\pi^2}\left(\frac{\pi a}{u_0 \beta}\right)^{2\pi K_0}\left( A_1(K_0)+\frac{ i  \beta \w}{2\pi}A_2(K_0)\right),
\end{equation}
where the dimensionless functions $A_1$ and $A_2$ are defined as
\begin{equation}
A_1(K_0)=\int_0^{\infty} \diff z \int_0^{\infty} \diff \bar{z} \, \frac{|z-\bar{z}|}{\left(\sinh z \sinh \bar{z}\right)^{\pi K_0}}, 
\qquad \qquad 
A_2(K_0)=\int_0^{\infty} \diff z \int_0^{\infty} \diff \bar{z} \, \frac{(z+\bar{z})|z-\bar{z}|}{\left(\sinh z \sinh \bar{z}\right)^{\pi K_0}}.
\end{equation}
For the memory function we find in the DC limit
\begin{equation}
M_{\rm{ps}}(T)=\frac{ i  u_0 K_0}{2a}y^2\left[\frac{\Gamma^4(\pi K_0/2)}{\Gamma^2(\pi K_0)} \left(\frac{2\pi a T}{u_0}\right)^{2\pi K_0-3}
\!\!-2^{3-2\pi K_0}\frac{1}{\pi}\sin\left(\pi^2 K_0\right)A_2(K_0)D_{Q}\left(\frac{2\pi a T}{u_0}\right)^{2\pi K_0-4} \right].
\label{Mps}
\end{equation}
A similar result for $D_{Q}=0$ was obtained in Ref. \onlinecite{Giamarchi91} for umklapp scattering in 1D systems.

In the opposite limit $D_{Q} u_0 \beta/a \gg 1$, contributions away from the diagonal $z=\bar{z}$ are suppressed. We thus find
\begin{align}
C_{\rm{ps}}(\w)&\approx \gamma \frac{u_0 \beta^2}{2\pi^2}\left(\frac{\pi a}{u_0 \beta}\right)^{2\pi K_0}\int_{0}^{\infty}\diff z \, \frac{e^{ i  \frac{\beta \w}{\pi} z}}{\left[\sinh\left(z \right)\right]^{2\pi K_0}} \int_{0}^{\infty} \diff \bar{z}\,\e^{-\frac{D_{Q}u_0\beta}{2\pi a}|z-\bar{z}|}
\\
&\approx \gamma \frac{2a^2}{u_0 D_{Q}} \left(\frac{2\pi a}{u_0 \beta}\right)^{2\pi K_0-1}\mathrm{B}\left(1-2\pi K_0,\pi K_0 -\frac{ i  \beta \w}{2\pi}\right).
\end{align}
In the zero frequency limit, the memory function reads
\begin{equation}
M_{\rm{ps}}(T)=\frac{2 i  u_0K_0}{a}\frac{\Gamma^2(\pi K_0)}{\Gamma(2\pi K_0)}\frac{y^2}{D_{Q}}\left(\frac{2\pi a T}{u_0}\right)^{2\pi K_0-2}.
\label{Mps1}
\end{equation}
\end{widetext} 
Equations (\ref{Mps}) and (\ref{Mps1}) yield Eq.~(\ref{Eq:CondpsShort-range}) of the main text.

\section{Long-range Coulomb interaction: RG analysis }
\label{App:RGLong}

In this Appendix we derive the RG equations describing the JJ chain with long-range Coulomb interaction, Sec.~\ref{Sub:RG-long}.
Our starting point is Eqs. (\ref{Eq:S0Final}) and (\ref{Eq:SpsQ}). 

\subsection{Lowest-order scaling\label{App:BareScaling}}

We start with the derivation of the scaling equations for $K$, $\ug$ and $y$ to first order in $y$.  
Straightforward dimensional analysis leads in this approximation  to RG equations (\ref{Eq:KLongZero}) and (\ref{Eq:ugLongZero}). 
To find the scaling of the QPS amplitude, we follow the standard route \cite{GiamarchiBook}  and perform averaging of the phase-slip part of the action over 
the eliminated modes. With the cutoff procedure described in the main text this leads to
\begin{equation}
\frac{d y(l)}{dl}=\frac{1+\ug}{2} y(l)\left[2-\pi K \epsilon\left(\ug\right)\right].
\end{equation}
Here
\begin{equation}
\epsilon(\ug)=\frac{4}{\pi(1+u_g)}\int_0^{1} d q \left(\frac{1}{q^2+1}
+\ug\frac{q^2\ug+1}{2q^2 \ug+1+q^2}
\right).
\end{equation}
The function $\epsilon(\ug)$ is smooth on the interval $0\leq\ug\leq 1$. Its value at $\ug=1$, $\epsilon(1)=1$, is universal and guarantees the correct scaling of the QPS amplitude at the infrared fixed-point $\ug=1$. 
On the other hand, the value of $\epsilon(\ug)$ at $\ug=0$ is non-universal and depends on the details of the cutoff procedure. Within our cutoff scheme  
$ \epsilon(0)=1$. Moreover, the full variation of  $\epsilon(\ug)$ on the interval $0\leq\ug\leq 1$ turns out to be numerically small (of the order of $1\%$).
We can thus safely assume $\epsilon(\ug)\equiv 1$ which leads us to Eq. (\ref{Eq:yLongRangeScaling}).

\subsection{Correlation functions and second-order correction\label{App:LongRangeSecondOrder}}

We are now in a position to derive the RG equations describing our system to the second order in $y$. 
To accomplish this goal we analyze the vertex function
\begin{equation}
R(\mathbf{r}_1)=\left<\e^{2  i  \phi(\mathbf{r}_1)}\e^{-2 i  \phi(0)}\right>.
\end{equation}
and examine its variation upon variation of the cutoff.  
The RG equations can then be read off from the requirement
\begin{equation}
\begin{split}
R_{\diff l}&(x_1(1+\diff l),\tau_1(1+\ug \diff l),\ug(0),K(0))
\\
&\hspace{2.5cm}=R_{l=0}(x_1,\tau_1,\ug(\diff l),K(\diff l)).
\end{split}
\label{A:Eq:RenormCond}
\end{equation}
Here $R_{\diff l}$ is the correlation function in the theory with the momentum cutoff $|q|\leq 1-dl$ while $R_{l=0}$ stands for the correlation function with initial cutoff $|q|\leq 1$.   

To zeroth order in $y$ 
\begin{equation}
R^{(0)}(\mathbf{r}_1)=\e^{-2 K F(\mathbf{r}_1)}
\end{equation}
with
\begin{equation}
F(\mathbf{r})=\pi^2\int_{|q|<1}\frac{\diff q}{2\pi}\int_{|\w|<\Omega_0}\frac{\diff \w}{2\pi}\, \,\frac{2-2\cos q x\cos \w \tau}{\frac{\w^2}{\Omega_0}+\frac{\Omega_0 q^2}{(1-\ug)q^2+\ug}}.
\label{Eq:Green-function}
\end{equation}
Equation (\ref{A:Eq:RenormCond}) leads then immediately to Eqs. (\ref{Eq:KLongZero}) and (\ref{Eq:ugLongZero}).
 
Let us now proceed with the perturbative treatment of the QPS action. 
In the   second order in $y$ we find a correction (cf. closely related discussion in Appendix \ref{App:RG-short_range})
\bea
\delta R &=& \frac{y^2 }{4\pi^3}K^2 \e^{-2K F(\mathbf{r}_1)} \int \diff^2 r \,\e^{-2 K F(\mathbf{r})}\e^{-2\pi^2 D_Q |x|} \nonumber
\\
& \times & \int \diff^2 R \left[\mathbf{r}\cdot \nabla_{\mathbf{R}}\left(F(\mathbf{R}-\mathbf{r}_1)-F(\mathbf{R})\right)\right]^2.
\label{Eq:second_order_correction}
\eea
Note that here the time is dimensionless (rescaled with $\Omega_0$). Since the function $F(\mathbf{r})$ is even in both $x$ and $\tau$, cross terms of the form $x\cdot \tau$ originating from the scalar product in the square bracket disappear. Transforming the integral over center of mass coordinates to Fourier space results in
\bea
\delta R &=& \frac{y^2 }{4\pi^3}K^2\e^{-2 K F(\mathbf{r}_1)}\int\frac{\diff^2q}{(2\pi)^2} \left[I_x q^2+I_{\tau}\w^2\right] \nonumber
\\
&\times & \left(2-2\cos\mathbf{q}\mathbf{r}_1\right) F^2(\mathbf{q}) \,,
\label{Eq:2ndOrderCorrectionForR}
\eea
where
\begin{equation}
I_\zeta=\int \diff^2 r \,\zeta^2 \,\e^{-2KF(\mathbf{r})}\e^{-D_Q |x|}, \qquad \zeta=x,\tau \,,
\end{equation}
and 
\begin{equation}
F(\mathbf{q})=-\frac{2 \pi^2}{\w^2+\frac{q^2}{(1-\ug)q^2+\ug}}.
\end{equation}

Evaluating the correction to $R$ after one step of RG, we get
\begin{widetext}
\bea
\delta R_{\diff l}(\tilde{x}_1,\tilde{\tau}_1) &=& \frac{y^2(0)}{4\pi^3}K^2(0) \e^{-2K(0) F_{\diff l}(\tilde{x}_1,\tilde{\tau}_1,\ug(0))} \int_{|q|<1-\diff l}\frac{\diff q}{2\pi}\int_{|\w|<1-\ug \diff l}\frac{\diff \w}{2\pi}\left(2-2\cos\mathbf{q}\tilde{\mathbf{r}}_1 \right) \nonumber
\\
&\times& F^2(\mathbf{q},\ug(0))\,\left[q^2\,I_{x,\diff l}(K(0),\ug(0),D_Q(0))  +\w^2\,I_{\tau,\diff l}(K(0),\ug(0),D_Q(0))\right].
\label{Eq:2ndOrderCorrectionOneStep}
\eea
\end{widetext}
We know from the zeroth order calculation that
\begin{eqnarray}
K(0) F_{\diff l}(\tilde{x}_1,\tilde{\tau}_1,\ug(0))&=&K^{(0)}(\diff l)F_0(x_1,\tau_1,\ug^{(0)}(\diff l)),\qquad\quad
\\x_1(1+\diff l)=\tilde{x}_1, &\quad& \tau_1(1+\ug \diff l)=\tilde{\tau}_1.\qquad\quad
\end{eqnarray}
The superscript $(0)$ emphasizes that only corrections for $y=0$ are taken into account at this stage. It is easy to show that
\begin{equation}
\begin{split}
&I_{x,\diff l}(K(0),\ug(0),D_Q(0))=\\&=(1+\diff l)^3 (1+\ug \diff l)\,I_{x,0}(K^{(0)}(\diff l),\ug^{(0)}(\diff l),D_Q(\diff l)),
\\
&I_{\tau,\diff l}(K(0),\ug(0),D_Q(0))=\\&=(1+\diff l) (1+\ug \diff l)^3 \,I_{\tau,0}(K^{(0)}(\diff l),\ug^{(0)}(\diff l),D_Q(\diff l)).
\end{split}
\end{equation}
The disorder strength is renormalized as $D_{Q}(\diff l)=(1+\diff l) D_Q$. We further need to rescale the integrals over $q$ and $\w$ in Eq. \eqref{Eq:2ndOrderCorrectionOneStep} as $\tilde{q}=(1+\diff l)q$ and $\tilde{\w}=(1+\ug\diff l)\w$. Further, we exploit
\begin{equation}
\begin{split}
&K^2(0)F_{\diff l}^2(\mathbf{q},\ug(0))=
\\
&=(1+\diff l)^2(1+\ug \diff l)^2\left(K^{(0)}(\diff l)\right)^2 F^2_0(\tilde{\mathbf{q}},\ug^{(0)}(\diff l)).
\end{split}
\end{equation}
Finally, we arrive at
\begin{widetext}
\begin{equation}
\begin{split}
\delta & R_{\diff l}(\tilde{r}_1,\ug(0),K(0),D_Q(0))=\frac{y^2(0)}{4\pi^3}(1+\diff l)^2(1+\ug \diff l)^2 \left(K^{(0)}(\diff l)\right)^2 \e^{-2K^{(0)}(\diff l) F_0(r_1,\ug^{(0)}(\diff l))}
\\
&\times \int_{|\mathbf{q}|<1}\frac{\diff^2 \tilde{q}}{(2\pi)^2}\left[\tilde{q}^2 I_{x,0}(K^{(0)}(\diff l),\ug^{(0)}(\diff l),D_Q(\diff l))+\tilde{\w}^2 I_{\tau,0}(K^{(0)}(\diff l),\ug^{(0)}(\diff l),D_Q(\diff l)\right]
\left[2-2\cos\tilde{\mathbf{q}}\mathbf{r}_1\right]F^2_0(\tilde{\mathbf{q}},\ug^{(0)}(\diff l)).
\end{split}
\label{Eq:2ndOrderCorrectionFinal}
\end{equation}
The full correlation function takes now the form
\bea
R_{\diff l}(\mathbf{r}_1) &=& \e^{-2 K F(\mathbf{r}_1)}\left\{1+\frac{y^2(\diff l)}{4\pi^3} K^2 \int \frac{\diff^2 q}{(2\pi)^2}\left[q^2 I_{x,0}+\w^2 I_{\tau,0}\right](2-2\cos\mathbf{q r}_1)F^2_0(\mathbf{q})\right\} \nonumber
\\
&\times& \biggl\{1+\frac{\pi}{2}(1+\ug)y^2(0)K^3 \diff l \int \frac{\diff^2 q}{(2\pi)^2}\left[q^2 I_{x,0}+\w^2 I_{\tau,0}\right](2-2\cos\mathbf{q r}_1)F^2_0(\mathbf{q})\biggr\} .
\label{Eq:2ndOrderCorrelatorDl}
\eea
\end{widetext}
Here,  we  have suppressed the superscript ${(0)}$ of $K$ and $\ug$ (in order to make the formula slightly less cumbersome) and used the rescaling law for the QPS amplitude,
\begin{equation}
y(\diff l)=\left[1+\left(1+\ug-\frac{\pi}{2}(1+\ug)K\right)\diff l\right]y(0).
\end{equation}
Finally, we need to compare Eq. \eqref{Eq:2ndOrderCorrelatorDl} to the correlator calculated at the original cutoff but with different couplings:
\begin{widetext}
\bea
R_{l=0} &= &\e^{-2 K(0) F_0(\mathbf{r}_1,\ug(0))}\left\{ 1-2 \,\delta K\, F_0(\mathbf{r}_1)-2 K \frac{\partial F_0}{\partial \ug}\delta \ug \right\}
\\ \nonumber
&\times& \left\{1+\frac{y^2(\diff l)}{4\pi^3} K^2 \int \frac{\diff^2 q}{(2\pi)^2}\left[q^2 I_{x,0} + \w^2 I_{\tau,0}\right](2-2\cos\mathbf{q r}_1)F^2_0(\mathbf{q})\right\}.
\eea
\end{widetext}
We introduced here the corrections as $\delta K=K(\diff l)-K(0)$ and $\delta \ug=\ug(\diff l)-\ug(0)$. 

If we would attempt to describe the renormalization of the quadratic action at all momenta $\bf q$, we would have to use a functional RG. Instead, we consider the long-wavelength limit of $F_0(\mathbf{q})$:
\begin{equation}
F_0(\mathbf{q})\simeq\frac{-2\pi^2}{\w^2+q^2/\ug}.
\end{equation}
This yields 
\begin{equation}
\begin{split}
\delta K&=-\frac{1}{2}(1+\ug) y^2 K^3 I_{\tau,0}\, \diff l,
\\
\delta \ug&= \frac{1}{2}(1+\ug) y^2 K^2 \ug (I_{\tau,0}-\ug I_{x,0})\, \diff l.
\end{split}
\end{equation}
This is sufficient both at the first stage of RG where the dominant effect in the renormalization of $K$ and $\ug$ is of zeroth order in $y$, as well as at longer length scales where the system approaches the local limit and our approximation will give the asymptotically correct form of the $y^2$ contributions to the renormalization.
Within our accuracy, the functions $I_{x,0}$ and $I_{\tau,0}$ can be evaluated in the local limit:
\begin{align}
I_x&\simeq C \frac{1}{K}\left[\BesselI_0(D_Q)-\StruveL_0(D_Q)-\frac{\BesselI_1(D_Q)-\StruveL_1(D_Q)}{D_Q}\right],
\\
I_{\tau}& \simeq C \frac{1}{K}\frac{\BesselI_1(D_Q)-\StruveL_1(D_Q)}{D_Q}.
\end{align}
Here $C$ is a numerical constant that we set to unity within our accuracy, and $\BesselI_n$ and $\StruveL_n$ are modified Bessel functions of the first kind and modified Struve functions, respectively. We are now in a position to write down the RG equations up to second order in the phase-slip fugacity:
\bea
\frac{\diff K}{\diff l}&=& -(1-\ug)K  \nonumber  \\ 
&-& \frac{1}{2}y^2 K^2 (1+\ug) \frac{\BesselI_1(D_Q)-\StruveL_1(D_Q)}{D_Q},  \\
\label{RG-long-range1}
\\
\frac{\diff y}{\diff l}&=&\frac{1+\ug}{2}\left[2-\pi K\right]y,   \label{RG-long-range2}
\\
\frac{\diff \ug}{\diff l}&=&2\ug(1-\ug)+\frac{y^2}{2}K(1+\ug)\ug   \nonumber
\\
&\times& \left[\!(1+\ug)\frac{\BesselI_1(D_Q)-\StruveL_1(D_Q)}{D_Q} \right. \nonumber \\
&-& \left. \ug \left(\BesselI_0(D_Q)-\StruveL_0(D_Q)\right)\!\right]\!,  
 \label{RG-long-range3}
\\
\frac{\diff D_{Q}}{\diff l}&=&D_{Q},  \label{RG-long-range4}
\eea
which leads to Eqs.~(\ref{Eq:yLongRangeScalingApprox})--(\ref{Eq:ugLongApprox}) of the main text.
Here the Bessel and Struve functions have the following asymptotic behavior:
\begin{align}
\BesselI_0(x)-\StruveL_0(x)&\sim 
\begin{cases}
1-\frac{2}{\pi}x, & x\to 0,
\\
\frac{2}{\pi x}, & x\to \infty,
\end{cases}
\\
\frac{1}{x}\left[ \BesselI_1(x)-\StruveL_1(x)\right]&\sim
\begin{cases}
\frac{1}{2}-\frac{3}{2\pi} x, & x\to 0,
\\
\frac{2}{\pi x}, & x\to \infty.
\end{cases}
\end{align}

Finally, we notice that, since we rescale frequencies as $\tilde{\w}=(1+\ug \diff l)\w$, temperature is also rescaled and satisfies the equation
\begin{equation}
\frac{\diff T}{\diff l}=\ug\, T.
\end{equation}
This is Eq.~(\ref{Eq:T-rescaling}) of the main text.

%%%%%%%%%%%%%%%%%%%%%%%%%%%%%%%%%%%%%%%%%%%%%%%%%%%%%%%%%%%%%%%%%%%%%%%%%%%%%%%%%%%%%%%%%%%%%%%%%%%%%%%%%%%%%%%%%%%%%%%%%%%%%%%%%%%%%%%%%%%%%%%%%%%%%%%%%
\section{Gaussian phase fluctuations: Comparison of JJ chains with superconducting nanowires}
\label{App:supercond-wires}
A model analogous to the one defined in Sec.~\ref{Sec:Model} is expected to describe also the physics of multichannel disordered superconducting wires at the SIT and deeply in superconducting and insulating phases\cite{MatveevEtAl02,Zaikin97}. Comparison of our action with that derived in Ref.~\onlinecite{Zaikin97} for the case of a dirty multichannel wire yields the following correspondence of parameters
\begin{equation}
\frac{1}{a E_0}\leftrightarrow\frac{\tilde{C}}{e^2}, \qquad \frac{a}{E_1}\leftrightarrow \frac{s\, \sigma}{e^2 \Delta}, \qquad a \Ej\leftrightarrow s\,\sigma \Delta/e^2,
\label{Eq:correspondences}
\end{equation}
where $\tilde{C}$ is the capacitance per unit length, $\sigma$ is the normal state conductance, $s$ is the cross section of the wire, and $\Delta$ is the modulus of the superconducting order parameter. We have also introduced the lattice spacing $a$ of the JJ chain into our action. The capacitance per unit length of the wire behaves as
\begin{equation}
\tilde{C}^{-1}\sim \ln(d/R),
\end{equation}
where $R$ is the radius of the wire and $d$ is the distance to a nearby metallic plate. Using the RHS of Eq.~\eqref{Eq:correspondences} to calculate the dimensionless parameter $K_0$ and the screening length $L_s\equiv a \Lambda$, we get
\begin{equation}
K_0 \sim \sqrt{\frac{\tilde{C}}{r_s}N_{\mathrm{ch}}}\,\frac{l}{\xi},
\qquad\qquad
L_s \sim \xi \sqrt{\frac{N_{\mathrm{ch}}r_s}{\tilde{C}}}.
\end{equation}
Here, $r_s\equiv e^2/v_F$ is the ratio of the interparticle spacing to the Bohr radius, $N_{\mathrm{ch}}$ is the number of channels in the wire, $l$ is the mean free path and $\xi$ is the coherence length of the superconductor in the dirty limit. 
While the model of a wire has a continuous character, the bare superconducting correlation length $\xi$ plays the role of the UV cutoff. Dividing $L_s$ by $\xi$, one can define an effective dimensionless screening parameter $\Lambda$
\begin{equation}
\Lambda \sim \sqrt{\frac{N_{\mathrm{ch}}r_s}{\tilde{C}}}.
\end{equation}
For a large number of channels, $\Lambda$ is much larger than unity. On the other hand, for a wire with just a few channels, $\Lambda$ can be of order unity.

The stray-charge disorder $D_Q$ has not been included in the model of Ref.~\onlinecite{Zaikin97}. Determination of its strength requires a separate analysis; we expect that the bare value of $D_Q$ becomes smaller with increasing $N_{\mathrm{ch}}$. Finally, while some results for the bare QPS fugacity $y$ were found in Ref.~\onlinecite{Zaikin97}, its analysis appears to require further work.

\end{appendix}


\begin{thebibliography}{100}
\bibitem{note-correlations}  To simplify the analysis, we restrict ourselves to the Gaussian fluctuations of $\xi$, which are  fully characterized by the second cumulant (\ref{Eq:Dxi}). On this level, there are no correlations between $\xi$ and $Q$. We do not expect any essential changes originating from the fluctuations beyond the Gaussian approximation.
\end{thebibliography}
\end{document}